\documentclass{aa}
\usepackage[T1]{fontenc}
\usepackage{subcaption}
\DeclareUnicodeCharacter{2014}{\dash}
\newcommand{\MOKA}{\ensuremath{\mathrm{MOKA^{3D}}}}
\usepackage{newunicodechar}
\usepackage{csquotes}
\usepackage{xcolor}

\newunicodechar{−}{-} 
\DeclareRobustCommand{\VAN}[3]{#2}
\let\VANthebibliography\thebibliography
\def\thebibliography{\DeclareRobustCommand{\VAN}[3]{##3}\VANthebibliography}

\usepackage{graphicx}	
\usepackage{amsmath}	
\usepackage[]{natbib}

\usepackage{txfonts}
\usepackage{hyperref}
\hypersetup{
    colorlinks = True,
    citecolor=blue,
    linkcolor = cyan,
}
\begin{document}

   \title{\MOKA: An innovative approach to 3D gas kinematic modelling}
    \subtitle{I. Application to AGN ionized outflows}

   \author{C. Marconcini
          \inst{1,2}
          \and
          A. Marconi
          \inst{1,2}
          \and
          G. Cresci
          \inst{2}        
          \and
          G. Venturi
          \inst{2,3,6}
          \and
          L. Ulivi
          \inst{1,2}
          \and
          F. Mannucci
          \inst{2}
          \and
          F. Belfiore
          \inst{2}
          \and
          G. Tozzi
          \inst{1,2}
          \and
          M. Ginolfi
          \inst{1,4}
          \and
          A. Marasco
          \inst{5}
          \and
          S. Carniani
          \inst{6}
          \and
          A. Amiri
          \inst{1,2,7}
          \and
          E. Di Teodoro
          \inst{1,2}
          \and
          M. Scialpi
          \inst{1,2}
          \and
          N. Tomicic
          \inst{1}
          \and
          M. Mingozzi
          \inst{8}
          \and
          M. Brazzini
          \inst{1,2}
          \and
          B. Moreschini
          \inst{1}
          }

   \institute{Dipartimento di Fisica e Astronomia, Università degli Studi di Firenze, Via G. Sansone 1,I-50019, Sesto Fiorentino, Firenze, Italy\\
              \email{cosimo.marconcini@unifi.it}
         \and
             INAF - Osservatorio Astrofisico di Arcetri, Largo E. Fermi 5, I-50125, Firenze, Italy\\
             \email{cosimo.marconcini@inaf.it}
         \and
             Instituto de Astrofísica, Facultad de Física, Pontificia Universidad Católica de Chile, Casilla 306, Santiago 22, Chile
        \and
             European Southern Observatory, Karl-Schwarzschild-Str. 2, D-85748 Garching, Germany
        \and 
             INAF, Padova Astronomical Observatory, Vicolo Osservatorio 5, 35122, Padova, Italy
         \and
             Scuola Normale Superiore, Piazza dei Cavalieri 7, I-56126 Pisa, Italy
        \and
             Department of Physics, University of Arkansas, 226 Physics Building, 825 West Dickson Street, Fayetteville, AR 72701, USA
        \and
             Space Telescope Science Institute, 3700 San Martin Drive, Baltimore, MD 21218, USA\\
             }

   \date{Received September 15, 1996; accepted July 4, 2023}

  \abstract
   {Studying the feedback process of Active Galactic Nuclei (AGN) requires characterising multiple kinematical components, such as rotating gas and stellar disks, outflows, inflows, and jets. To compare the observed properties with theoretical predictions of galaxy evolution and feedback models and to assess the mutual interaction and energy injection rate into the interstellar medium (ISM), one usually relies on simplified kinematic models.
   These models have several limitations, as they often do not take into account projection effects, beam smearing and the surface brightness distribution of the emitting medium.
   Here, we present \MOKA, an innovative approach to model the 3D gas kinematics from integral field spectroscopy observations. In this first paper, we discuss its application to the case of AGN ionised outflows, whose observed clumpy emission and apparently irregular kinematics are only marginally accounted for by existing kinematical models.
   Unlike previous works, our model does not assume the surface brightness distribution of the gas, but exploits a novel procedure to derive it from the observations by reconstructing the 3D distribution of emitting clouds and providing accurate estimates of the spatially resolved outflow physical properties (e.g. mass rate, kinetic energy).
   As an example, we demonstrate the capabilities of our method by applying it to three nearby Seyfert-II galaxies observed with MUSE at the VLT and selected from the MAGNUM survey, showing that the complex kinematic features observed can be described by a conical outflow with a constant radial velocity field and a clumpy distribution of clouds.}


   \maketitle
\section{Introduction}
Feedback models for galaxy evolution are mostly based on theoretical predictions, hydrodynamical simulations, and kinematical models based on observations \citep{fabian2012, morganti2017, Kudritzki2021}. Therefore, an in-depth comprehension of multi-phase gas kinematics in galaxies is crucial to determine the mutual interaction of ISM phases and shed light on energy injection and ejection mechanisms. 
Integral Field Unit (IFU) observations, together with kinematical models, allow to constrain with unprecedented detail and accuracy the 2D and 3D kinematics of gas disks in galaxies \citep{oh2008, diteo2015, boiche2015}.
Despite the considerable advances in the modelling of gas kinematics in galaxy disks, galaxy-wide multi-phase outflows are still poorly constrained, due to their complex observed geometry, kinematics and surface brightness distribution \citep[e.g.][]{Veilleux2020}.\par
The unified model of AGN predicts that radiation from the accretion disk around the supermassive black hole is collimated by a torus of obscuring dust and gas, which is broadly symmetric around the accretion flow axis \citep{UrryPadovani1995}. This radiation impinges on the ISM, ionising the gas and transferring mechanical energy through radiation pressure and disk winds. Consequently, the outflowing gas extends outward from the nucleus assuming a typical (bi)conical geometry. These cones are more frequently observed in local Seyfert galaxies, especially in type-II (obscured) than type-I (unobscured) Seyferts, consistently with the unification model \citep{Schmitt2016}.
The best tracer of the ionized phase of these outflows on $10^{2}-10^{3}$ pc scales is the forbidden emission line doublet [OIII]$\lambda\lambda4959,5007$, since it can only be produced in low density regions, thus not tracing the sub-parsec scales of the Broad Line Region (BLR). In the presence of outflows, the spectral profile of the [OIII]$\lambda5007$ emission line typically assumes an asymmetric shape, with a broad, blue-shifted wing\footnote{A similar reasoning applies for the [NII]$\lambda6584$ emission line, a secondary useful tracer of the ionized emission in those sources, where a possible [OIII]$\lambda5007$ blue wing may be undetected due to dust obscuration.} \citep[e.g.][]{Boroson2005, Zakamska2014, Perna2017, Bae2018}.
A detailed study of gas kinematics and ionisation mechanisms in these cones is extremely important to shed light on the interaction between AGN and their host galaxies, and on the multi-phase nature of outflows \citep[e.g.][]{Crenshaw2015, Karouzos2016, Bae2018, Cresci2018, Venturi2018, Mingozzi2018, Fluetsch2019, Kraemer2020, Marasco2020, Tozzi2021}.
AGN-driven outflows represent the main manifestation of radiative feedback in action, having the potential to sweep away the gas content of the host galaxy, thus quenching star formation inside the outflow cavity from low to high redshift \citep[e.g.][]{Feruglio2010, Hopkins2010, Sturm2011, CanoDiaz2012, Zubovas2012, Cicone2014, Carniani2015, Cresci2015a, Carniani2016}. As a result, they can shape galaxies and cause the well-known empirical relations between the black hole mass ($M_{\rm BH}$) and host galaxy properties, such as mass or luminosity of the bulge ($M_{\rm bulge}$ and $L_{\rm bulge}$, respectively) and velocity dispersion ($\sigma$) \citep{Gebhardt2000, Marconi2003, Ferrarese2005}.\par
Even though AGN-powered outflows are thought to play a key role during galaxy evolution, an accurate determination of outflow properties and driving mechanisms is still missing, due to the large uncertainties related with the estimated outflow properties in different gas phases \citep{Cicone2014, veilleux2017, harrison2018}. In most sources, due to unresolved observations, it is unclear whether the outflow morphology is conical or shell-like, and it is largely debated what physical processes drive the coupling of the energy and of the momentum released by the AGN with the surrounding ISM. 
The most accepted model predicts that a nuclear highly ionized wind, accelerated by the BH radiation pressure, shocks the ISM of the host galaxy, creating an expanding bubble of hot gas, with the post-shock medium going through a momentum-conserving phase, followed by an adiabatically expanding phase \citep{King2003, King2011, Cicone2014, Feruglio2015, Tombesi2015}.
To address these debated topics, many kinematical models have been proposed to analyse observations of galactic outflows, but up to date just a few have been tested with spatially resolved integral field spectroscopy (IFS) data (e.g. the AGN-outflow model of VLT/SINFONI data from \citet{MullerSanchez2011}). Simpler models based on long-slit observations have been proposed to investigate the kinematics of the Narrow Line Region (NLR) in local Seyfert galaxies, modelling the position–velocity (PV) diagram \citep{Crenshaw2000a, Crenshaw2000b, Das2005, Fischer2010, Storchi2010}.\par
\citet{Venturi2017} made a first comparison between the observed moment maps obtained from MUSE data of NGC 4945 and NGC 1365, and a kinematic toy model to investigate the outflow kinematics. This toy model is the starting point for our current kinematic model.
Up to now, literature models, such as those mentioned above, are computed by adopting a conical outflow geometry, with velocity field and gas surface brightness of the emitting gas parameterized by analytical functions of the distance from the cone vertex. 
The velocity profile, at a given position on the sky is given by:
\begin{equation}
    f_P(v) = \int_{LOS} \Sigma_P(s,v) ds
\end{equation}\label{eq.1}\noindent
where $P$ is a given position on the sky, $\Sigma_P(s,v)$ is the corresponding gas surface brightness distribution at a given velocity $v$ and coordinate $s$ along the line of sight (LOS). 
The best parameters are then found to reproduce $ f_P(v)$ for any given spaxel, or its average velocity and velocity dispersion.
In few cases the model takes into account the finite spatial and spectral resolution by convolving the data with the proper smoothing kernels \citep[e.g.][]{Bae2016, Karouzos2016}.
However, the assumption of a smooth function $\Sigma(s,v)$ to describe the gas emissivity is in contrast with observations, which in nearby galaxies show clumpy and irregular structures.
Therefore, an important question to be addressed is whether the complex velocity field observed in nearby sources - and never reproduced by previous studies - is the result of a complex velocity distribution within the propagating outflow, or the consequence of a clumpy distribution of ionized gas clouds.\par
To address these debated questions, we propose an innovative approach to model gas kinematics and determine outflow properties, explaining improvements with respect to previous methods.
In this paper, we present our new kinematic model  \MOKA\ (Modelling Outflows and Kinematics of Agn in 3D), discuss its main degeneracies and show examples of its application to three Seyfert II galaxies from the MAGNUM survey (Measuring Active Galactic Nuclei Under MUSE Microscope, \citealt{Cresci2015b, Venturi2018, Mingozzi2018, Venturi2021}), NGC 4945, Circinus, and NGC 7582.\par
The paper is constructed as follows. In Sect. \ref{kin_mod_chapter}, we describe step by step the 3D model operation. In Sect. \ref{mock_tests} we present the application of our method to simulated data. In Sect. \ref{magnum_test}, we introduce the sample to which we applied the model, describing the gas kinematic features, and the results for the kinematics and energetics obtained following the model application. In Sect. \ref{conclusions_section}, we summarize our results and present future developments. 
\section{Kinematic model}\label{kin_mod_chapter}
\begin{figure}
    \includegraphics[width=0.95\linewidth]{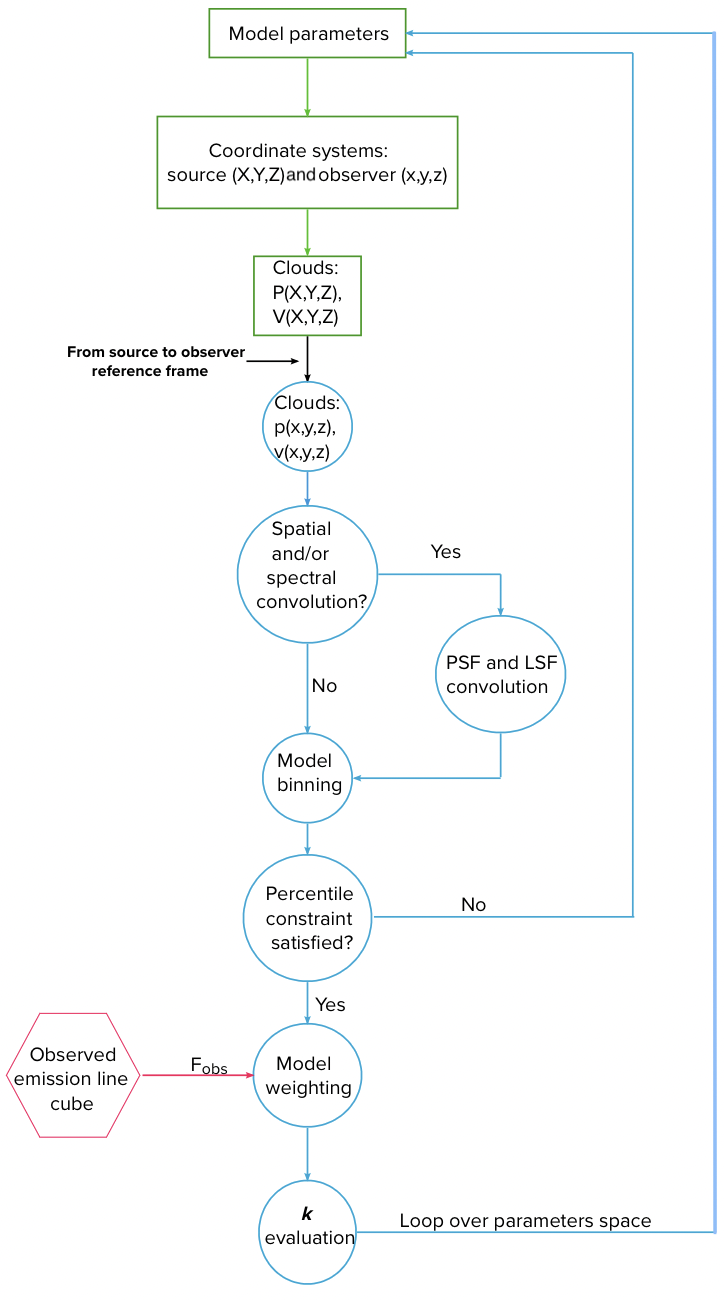}
    \caption{Flowchart of our \MOKA \ kinematic model. The method consists in assigning the model parameters and assuming a coordinate system and a velocity field. Each cloud is assigned the coordinates in the source reference frame (green), which is later transformed to the observer reference frame (blue), through Euler rotation matrices. If necessary, the methods requires the convolution with the PSF and/or the spectral line spread function (LSF). Then, the model is binned in the space sampled by the observed data cube (red). The model emission line profile percentiles at 1\% and 99\% are compared to the observed ones and, if they are comparable, each particle is weighted according to the observed emission in the corresponding bin.
    The model is evaluated by means of the $\kappa$ estimator (Eq. \ref{kappa_goodness}). The free parameters are varied and the loop repeats, until finding the suitable parameter configuration that minimizes $\kappa$ and reproduces the observed spectrum in each spaxel.
    \label{model_scheme}}
\end{figure}

In this section we present an overview of our model, describing the main steps to infer the kinematics and orientation of observed galactic structures.
In Fig. \ref{model_scheme}, we show a schematic flowchart of the \MOKA \ model operation and fitting, described in the following.

\subsection{Model set up}
We adopt a spherical coordinate system and create a uniform distribution of 10$^7$ point-like synthetic emitting sources (clouds, hereafter) distributed according to an input geometrical distribution chosen by the user (e.g. a cone when modeling outflows). We decide to adopt this number of clouds after considering the average MUSE emission line data cube size, that is $\sim$ 300$\times$300$\times$40. In particular, as discussed in Sect. \ref{unweighted_model}, we need at least a few model cloud per voxel (volume pixel) to properly model the observed features.  The spherical coordinate system is selected to model radial outflows, but the method works with any kind of reference system, such as cylindrical or Cartesian. The 3D spatial location of each cloud is specified by three coordinates: the semi-polar angle $\theta$, measured from the cone axis ($0^{\circ} \le \theta \le 180^{\circ}$); the azimuthal angle $\phi$ of the cloud orthogonal projection on a reference plane passing through the origin and orthogonal to the axis (measured clockwise, $0^{\circ} \le \phi \le 360^{\circ}$); and the radial distance $r$ from the origin (in arcsec). 
We start by assuming the surface brightness distribution of the emitting clouds as follows:
\begin{equation}
        f(r) = f_{\rm 0} \ e^{-\frac{r}{r_{\rm 0}}},
    \label{fluxexpo}
\end{equation}
\noindent
where $f_{\rm 0}$ is the flux value at the apex of the cone and $r_{\rm 0}$ is an arbitrary scale-radius.
For what concerns the outflow velocity, we start by assuming it as constant with radius ($\rm V(r) = V_{\rm 0}$).
\subsection{Reference frame}
\begin{figure}
    \includegraphics[width=\linewidth]{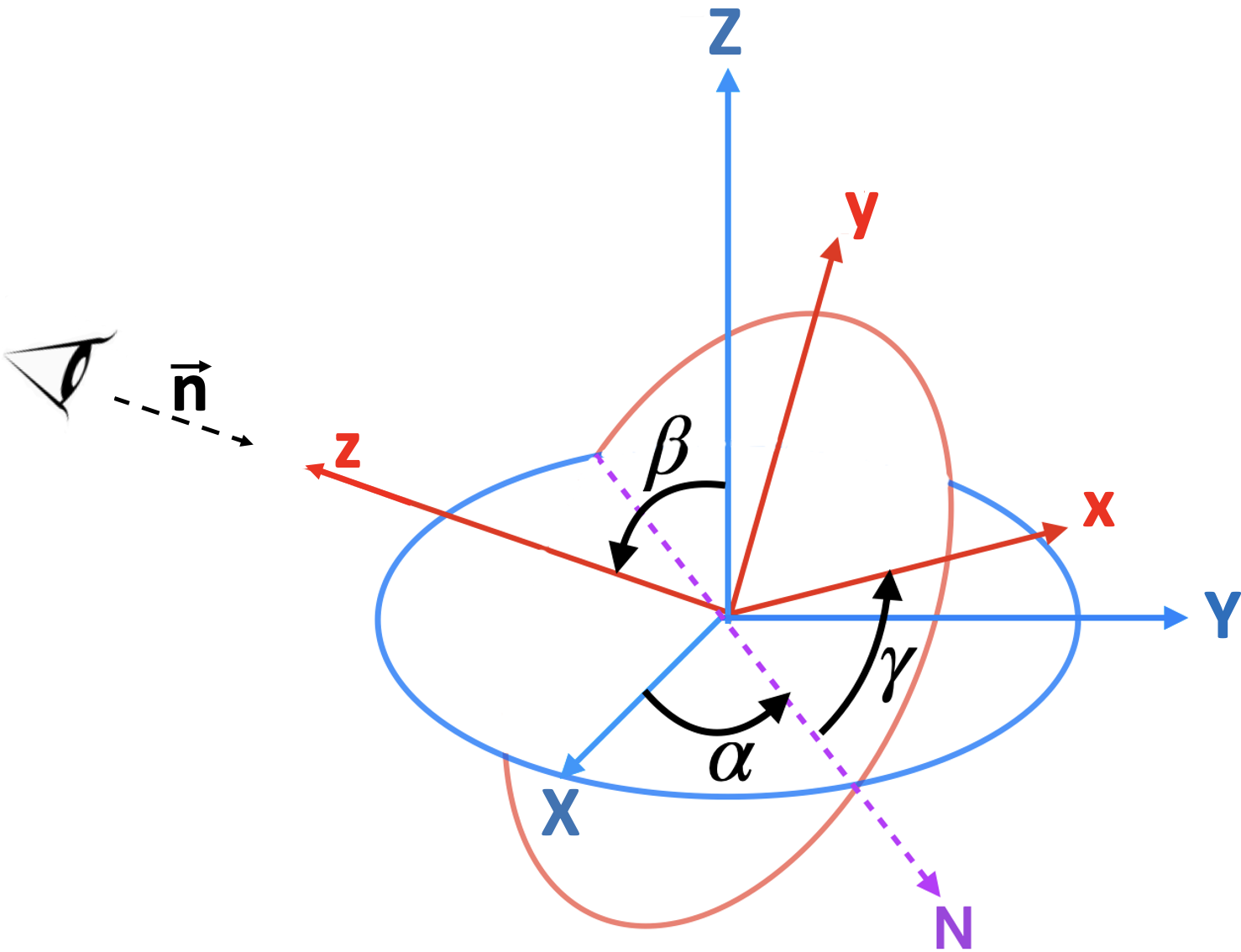}
    \caption{Source (blue) and observer (red) reference frame. Euler angles $\alpha$, $\beta$ and $\gamma$ transforms from the source to the observer's frame and vice-versa. Purple dotted line represents the line of nodes, N. The observer $z$ axis coincides with LOS, ($x, y$) is the plane of the sky.
    \label{eulero}}
\end{figure}

We define the source reference system in cartesian coordinates as $\rm (X, Y, Z) = (\sin(\theta)\cos(\phi),  \sin(\theta)\sin(\phi), \cos(\theta))$. Therefore, the status of each cloud is specified by position and velocity vectors, namely, $\rm \vec{P}^{\,} = (X, Y, Z)$ and $\rm \vec{V}^{\,} = (\vec{V_X}^{\,}, \vec{V_Y}^{\,}, \vec{V_Z}^{\,})$, with the velocity vector defined as:
\begin{equation}
    \rm \vec{V}^{\,} = v_{\rm 0}  \vec{u_r}^{\,} = v_{\rm 0} (\hat u_X, \hat u_Y, \hat u_Z) = v_{\rm 0} \begin{bmatrix}\sin(\theta)\cos(\phi) \\ \sin(\theta)\sin(\phi) \\ \cos(\theta) \end{bmatrix} 
\end{equation}
\noindent
It is also possible to associate to each cloud a random velocity dispersion component $\vec{\sigma}^{\,}_{\rm rand}$, but in this work we have not taken advantage of this possibility.
Then, we consider the observer's frame $(x,y,z)$, where $(x,y)$ are the coordinates on the plane of the sky and $z$ is directed along the LOS. The source and observer reference frame are shown in Fig. \ref{eulero}, as blue and red, respectively.
We transform $\vec{P}^{\,}$ and $\vec{V}^{\,}$ from the source to the reference frame by means of Euler rotation matrices:
\begin{equation}
    \begin{bmatrix}
           x \\
           y \\
           z
         \end{bmatrix} = \rm R_{\gamma} R_{\beta} R_{\alpha} \begin{bmatrix}X \\ Y \\ Z \end{bmatrix}
\end{equation}\noindent
where $\rm R_{\gamma}, R_{\beta}, R_{\alpha}$ are the Euler rotation matrices, with the rotation angles shown in Fig. \ref{eulero} and defined as:
\begin{itemize}
    \item $\alpha$: Corresponds to the angular separation between the line of nodes and the right ascension coordinate in the observer reference system. The $\alpha$ angle is irrelevant if the observed source is axially symmetric around the source z-axis.
    \item $\beta$: Corresponds to the outflow axis inclination with respect to the plane of the sky (e.g. $\beta = 180^{\circ}$ in case of a source pointing away from the observer, with z-axis along the LOS; $\beta = 90^{\circ}$ for a source axis lying on the plane of the sky).
    \item $\gamma$: Corresponds to the projected source major axis inclination in the plane of the sky with respect to the line of nodes, indicating the rotation direction. Also known as Position Angle (P.A.).\\
\end{itemize}
\hfill

The velocity component along the LOS, in the observer reference frame, is defined as: 

\begin{equation}
    \rm v_z =  \vec{V}^{\,} \cdot  \vec{n}^{\,} + v_{sys} = \vec{V}^{\,} \cdot \begin{bmatrix}0 \\ 0 \\ -1 \end{bmatrix}  + v_{sys}
    \label{v_los_def}
\end{equation}
\noindent
with $\vec{n}^{\,}$ the unit vector pointing away from the observer, $v_{sys}$ the outflow systemic velocity with respect to the galaxy. 
Once the clouds are projected, the model allows for both a spatial and spectral convolution, to account for observational effects. The observed position $(x,y)$ of each cloud is randomly shifted on the plane of the sky according to a 2D probability distribution, given by the point spread function (PSF) measured from the data.
If necessary, to account for the intrinsic spectral resolution, the observed LOS velocity can be similarly convolved with the line spread function (LSF), but in this work we have not taken advantage of this possibility.
In the following, we will refer to the outcome of this step as the unweighted model.
\subsection{Unweighted model}\label{unweighted_model}
\begin{figure}
    \includegraphics[width=\linewidth]{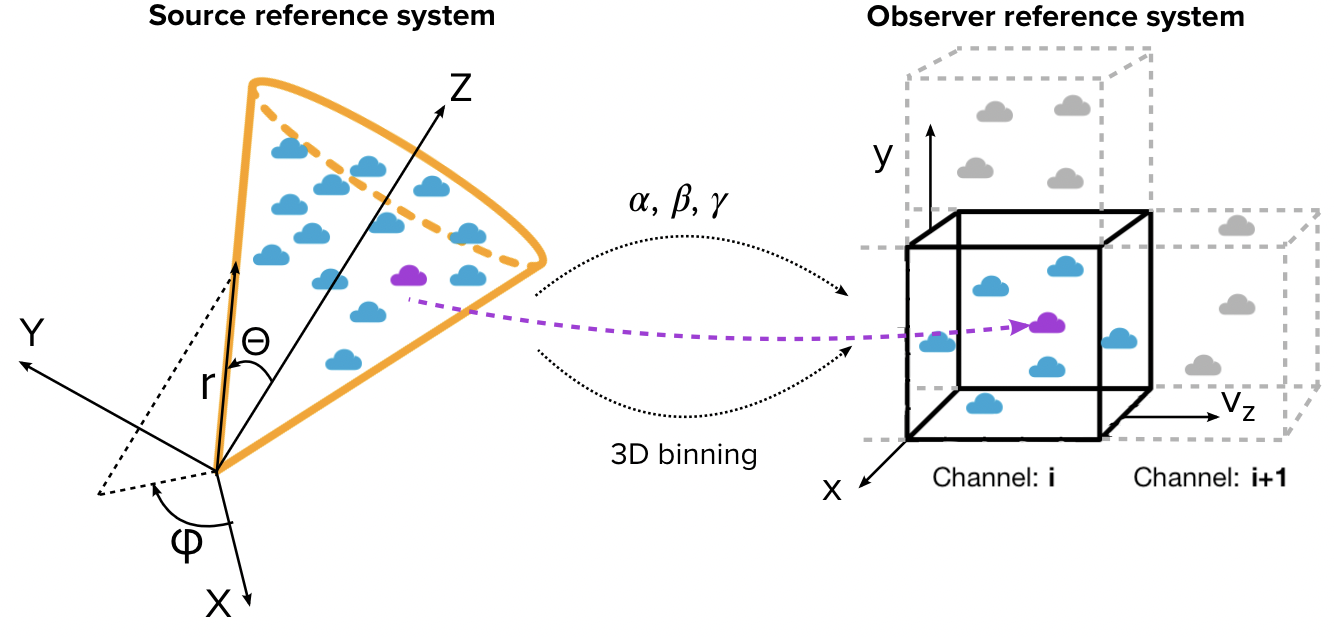}
    \caption{Coordinate transformation and binning procedure of the model: from the simple conical geometry in the source reference frame, uniformly populated with clouds, to the 3D model cube in the observer's reference frame. The model is then binned to assign to each cloud a position in the $(x\ , y\ , v_{\rm z})$ space, as highlighted for the purple cloud. The model cube has the same spatial and spectral resolution as the observed data cube, to weight the clouds according to the observed flux in the corresponding spaxel. Dashed grey lines show two adjacent voxels, corresponding to different spatial and/or spectral channels.
    \label{figura4.00}}
\end{figure}

At this stage the unweighted model is binned in the observed $(x,y,v_z)$ space, which is the one sampled in data cubes produced by IFU observations. The process is schematized in Fig. \ref{figura4.00}. The spatial and spectral extensions of the model cube are forced to match the spatial and spectral sampling of the data, respectively. A voxel of the obtained 3D model cube, with two spatial extensions and a spectral-velocity one, uniformly populated with clouds, is highlighted in black in Fig. \ref{figura4.00}.
In this phase, all clouds in the same bin have a weight associated to our initially assumed radial-dependent flux function (Eq. \ref{fluxexpo}). In particular, all the clouds corresponding to the same spaxel have the same intensity, being dependent only on the radial distance from the vertex of the cone. Spreading homogeneously 10$^7$ clouds in the binned model cube guarantees that at least few clouds populate each voxel, and thus that the weighting procedure described in the following Sect. \ref{weighted_model} succeeds. Since multiple clouds can fall in the same bin, as shown in Fig. \ref{figura4.00}, their contribution to the emission of the corresponding model spectral channel is assumed to be equal.\par
\begin{figure}
    \includegraphics[width=\linewidth]{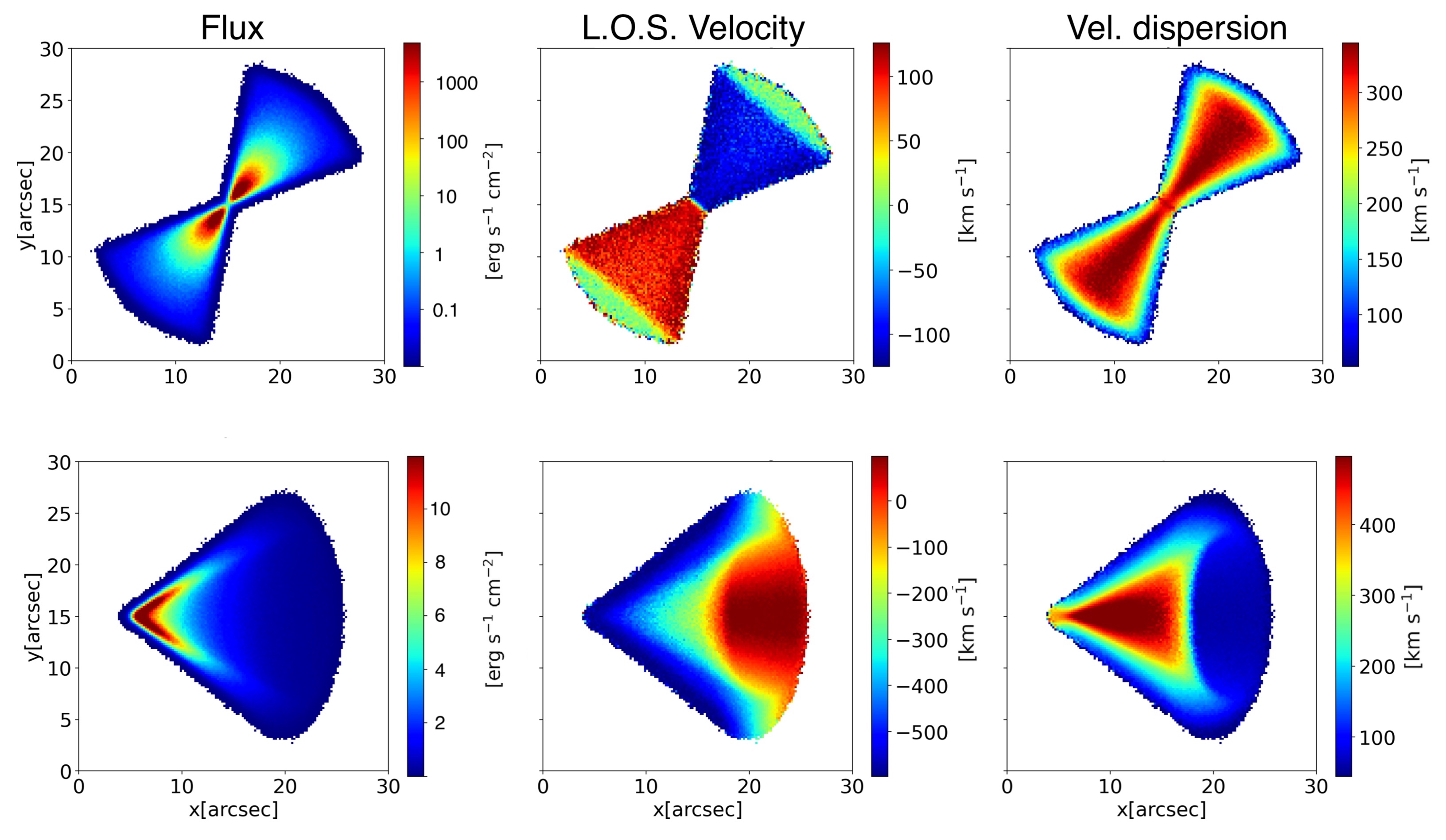}
    \caption{Example of unweighted models simulating an outflow with spatial and spectral resolution of a galaxy observed with MUSE at $z=0.001$. From left to right: Line-velocity integrated map, LOS velocity map and velocity dispersion map. $Top \ Panels$: Full biconical model with $P.A.=40^{\circ}$, inclination with respect to the plane of the sky of $\beta = 85^{\circ}$ and constant radial velocity of $1200$ km s$^{-1}$, the cone semi-aperture is $\rm \theta_{OUT} = 30^{\circ}$. $Bottom \ Panels$: Single hollow conical model with $P.A.=90^{\circ}$, $\beta = 65^{\circ}$, inner and outer semi-aperture $\rm \theta_{IN} = 25^{\circ}, \theta_{OUT} = 35^{\circ}$.  Both models have been convolved with a spatial PSF = 0.3''and no spectral convolution. 
    \label{figura4.2}}
\end{figure}

Once the unweighted model is computed, each spaxel (spectral pixel) has an associated spectrum (see Eq. \ref{eq.1}), which depends on the model geometry and the parameters of the flux and velocity functions we assumed. This model spectrum can be compared with the corresponding observed one in the same spaxel.
From the model cube it is then possible to compute projected flux, velocity and velocity dispersion maps from the momenta of the cube.
In Fig. \ref{figura4.2} we show two examples of unweighted models: a filled bi-conical outflow (top panels) and a hollow conical one (bottom panels). 

\subsection{Weighted model}\label{weighted_model}
We use the line flux $\rm F_{\rm obs}$ from each  $(x\ , y\ , v_z)$ voxel of the observed data cube, to determine the surface brightness of each model cloud.
In particular, we assign a weight $\rm w = F_{\rm obs}/N_{\rm mod}$ to each cloud within the volume element specified by the $(x\ , y\ , v_z)$ coordinates, where $\rm N_{\rm mod}$ is the total number of clouds ending up in that voxel. We want to clarify that, hereafter each cloud is assigned a weight which is independent of the distance from the vertex of the cone. Therefore, each cloud emission is not related to previous assumptions on the analytical surface brightness distribution (this distribution could be any function, see e.g. Eq. \ref{fluxexpo}). As an example, all the clouds in the highlighted black voxel in Fig. \ref{figura4.00} have the same weight, since they belong to the same voxel and thus are weighted with the same flux.
We can now compare the observed line profile in each spaxel with those obtained from a variety of weighted model cubes, each one computed using different combinations of geometrical and physical parameters.
Assigning a weight to each cloud ensure that the model velocity profile will match the observed one by construction, provided that model clouds occupy the same velocity range spanned by the data. The latter condition will not be satisfied if the model geometry and velocity field is incorrect, which gives us the means to constrain both the cone geometry and the intrinsci velocity field at the same time.
The procedure to infer the outflow parameters through a fitting procedure is described in the next section.

\subsection{Fit of model on observed data}\label{fit_subs}
The fitting algorithm consists in a loop over the parameter space, until the best set of parameters is obtained, which is that one minimizing the difference between the observed and modelled emission in each voxel. First, we define the free parameters we want to explore for the MUSE example: intrinsic radial outflow velocity, outflow systemic velocity with respect to the galaxy systemic velocity ($\vec{V}^{\,}$ and $\rm v_{sys}$ in Eq. \ref{v_los_def}, respectively), and inclination with respect to the LOS. Then, following the procedure shown in Fig. \ref{model_scheme}, for each set of parameters we create a weighted model cube, whose emission is compared to the observed one over the velocity range defined by the 1$\%$ and 99$\%$ percentiles of the observed LOS velocity distribution, $\rm v_{1}$ and $\rm v_{99}$, respectively. Finally, we evaluate how well each model reproduces the observed spectrum by means of a customized goodness of fit estimator, defined as:

\begin{equation}
    \kappa = \sum_{i,j} \left(\frac{S_{\rm Oj}(v_i) - S_{\rm Mj}(v_i)}{\delta s_j} \right)^2 \label{kappa_goodness}
\end{equation}
\noindent
Here, $v_i$ represent the i-th spectral channel of the j-th spaxel.
Therefore, $S_{\rm Oj}(v_i)$ and $S_{\rm Mj}(v_i)$ are the observed and model spectral flux in the i-th voxel, respectively, and $\delta s_j$ is the uncertainty on the flux, assumed to be constant in all spectral channels.
We minimize $\kappa$ to find the best set of model parameters.
Although the $\kappa$ estimator is defined as the standard $\chi^2$ estimator, its definition relies on different assumptions. For this reason, we decided to refer it as $\kappa$ instead of $\chi^2$. A proper computation of the free parameters uncertainties and the development of a tailored statistic will be addressed in future work.
\begin{figure}
    \includegraphics[width=\linewidth]{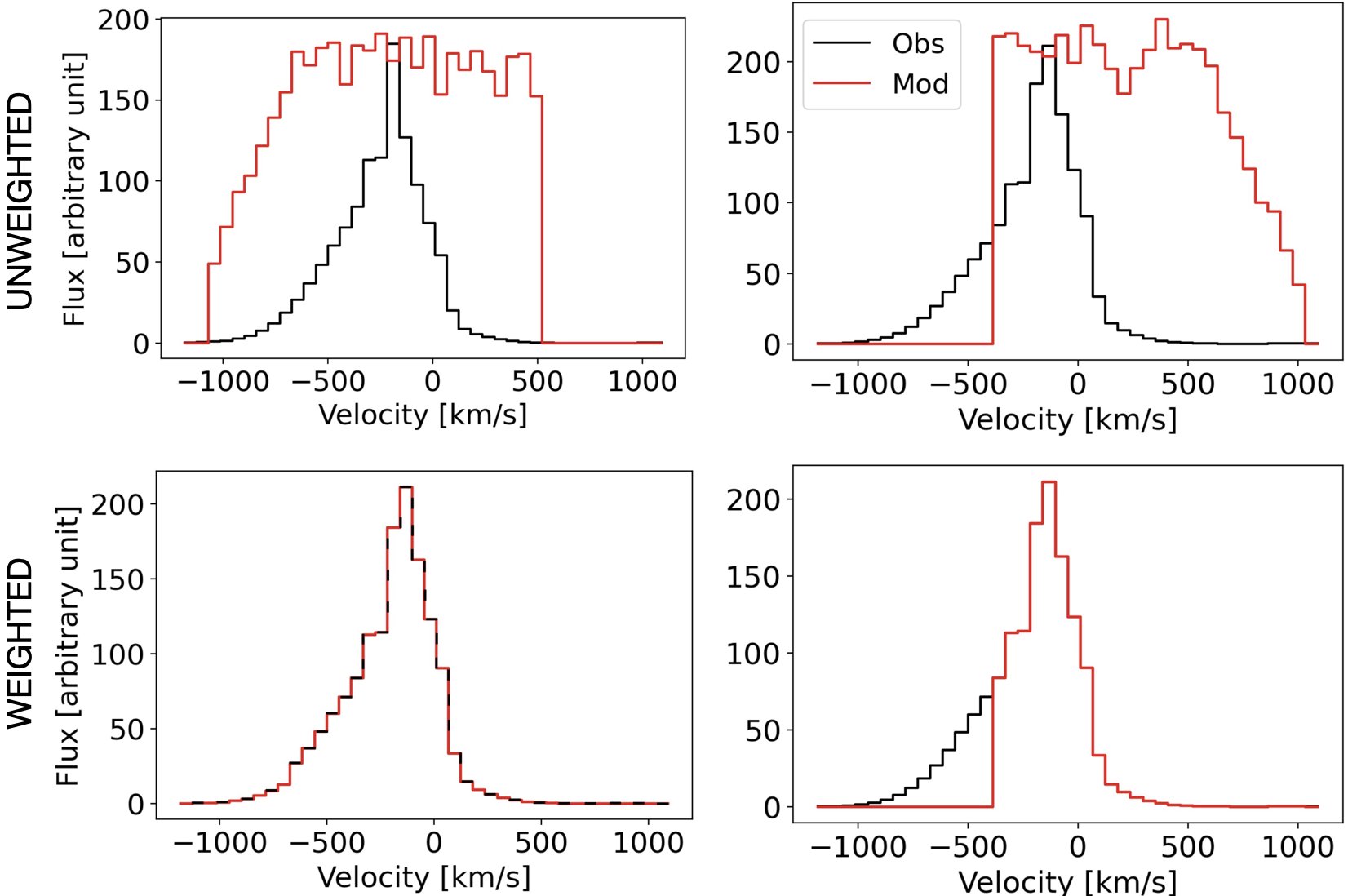}
    \caption{Comparison between the observed [NII]$\lambda$6584 emission line profile (black) and the model emission (red) of NGC 4945, extracted from a spaxel with a prominent blue-shifted wing. 
    Top and bottom panels show the unweighted and weighted emission, respectively. Left panels: the model emission is obtained with the best fit parameters reported in Table \ref{tab.1}. Right panels: as a demonstrative example, we show the model line profile as obtained by assuming an unsuitable inclination and opening angle. Such a configuration is clearly not able to to reproduce the high-velocity, blue-shifted emission, thus causing a sharp cut off. The absence of clouds in all velocity channels where emission is observed, prevent the model to reproduce the observed outflow properties.
    \label{line_prof_comparison}}
\end{figure}
Figure \ref{line_prof_comparison} shows the comparison between the observed (black) and model (red) [NII]$\lambda$6584 emission line profile, extracted from a spaxel with a prominent blue-shifted wing in NGC 4945. Top panels show the unweighted model emission for the best fit model (left) and a model with wrong inclination (right). Bottom panels show the weighted version of top panels.
The best model configuration on the bottom left is obtained with the parameters listed in Table \ref{tab.1}, and perfectly reproduces the observed emission in each velocity channel.  
In the right panels, a different set of model parameters is not able to reproduce the observed blue wing, for either the weighted or unweighted model. This is due to the fact that, for this wrong set of parameters, no cloud contributes to observed LOS velocities smaller than $-300$ km s$^{-1}$. Indeed, for a model to be successful, it is crucial that in all spaxels there are model clouds populating all velocity channels where emission is observed. 
\subsection{Solving the degeneracies}\label{degeneracies}
The main problem to face when creating a 3D kinematical model is recognizing the degeneracies which affect the fit results. 
The main degeneracy affecting \MOKA \ is that very high outflow velocities, combined with a range of different inclinations towards the observer's LOS, always allow to reproduce the observed line profile in each spaxel, even with a wrong set of geometrical parameters, since all velocity channels of each spaxel will be always populated with model clouds.\par
\begin{figure}
    \centering
    \includegraphics[width=\linewidth]{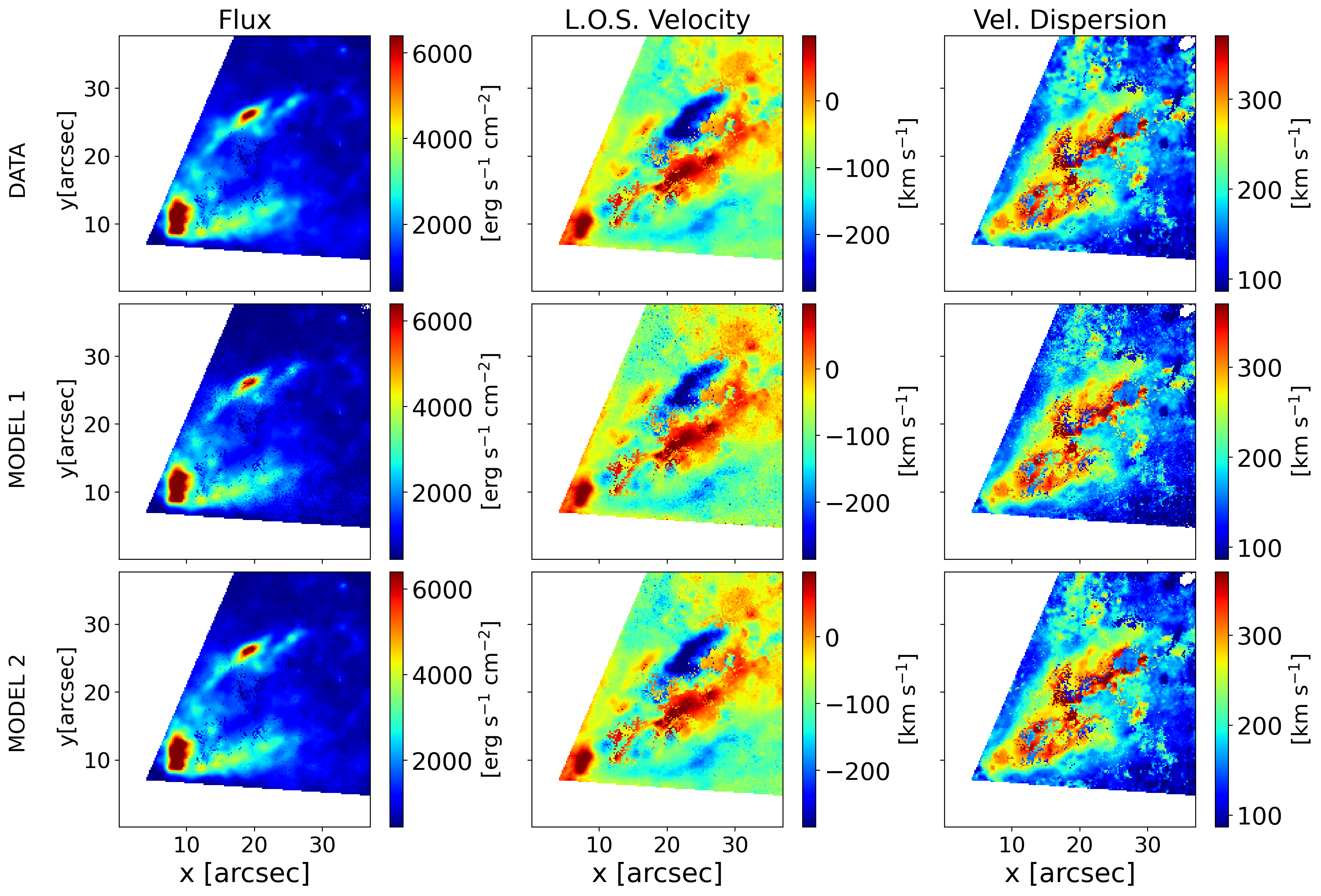}  
    \caption{Degenerate models of NGC 4945: moment maps of observed data (top panels) and of two models with high radial outflow velocity and wrong inclination with respect to the plane of the sky, outer opening angle and systemic velocity (middle and bottom panels), with respect to the best model shown in Fig. \ref{figura4.4}. See Sect. \ref{degeneracies} for details of degenerate models' parameters and how to overcome them.}
    \label{degeneracy_maps}
\end{figure}

As an example of this, Fig. \ref{degeneracy_maps} shows the moment maps of observed data (top panels), and two models with radial outflow velocity of 3000 km s$^{-1}$, which is above the observed maximum velocity of $\sim$ 900 km s$^{-1}$. The model in middle panels is obtained with $\beta$ = 50$^{\circ}$, v$_{\rm sys}$ = +1000 km s$^{-1}$ and outer semi-opening angle of $\theta_{\rm out} = 48^{\circ}$; the model in bottom panels instead, with 
$\beta$ = 120$^{\circ}$, v$_{\rm sys}$ = -500 km s$^{-1}$ and outer semi-opening angle of $\theta_{\rm out} = 38^{\circ}$.\par
Figure \ref{degeneracy_maps} shows that models with high radial outflow velocity are able to equally reproduce the observed emission and kinematics in each spaxel, even with a very different geometrical configuration.
If the outflow cone aperture is wide enough to reproduce the observed cone aperture in the plane of the sky, the combination of high velocities and a wide range of inclination angles with respect to the plane of the sky, can provide a wide range of velocities along the LOS, fitting any kind of observed data, once the weighting scheme described above is applied. \par
Since each cloud is assigned a weight, measured from the observed data, and dependent on the $(x, y, v_z)$ bin where the cloud falls, there are two configurations causing a cloud to be assigned a zero weight: either the cloud has a velocity $v_z$ beyond the observed velocity ranges, or no emission is detected in the cloud 3D position-velocity. 
Therefore an outflow radial velocity which results in velocity percentiles well above the observed boundaries $\rm v_{1}$ and $\rm v_{99}$, as the cases shown in Fig. \ref{degeneracy_maps}, allows the model to reproduce the observed spectrum in each spaxel by assigning zero weight to clouds at the model edges, thus producing a flattened model on the plane of the sky.\par
\begin{figure}
    \centering
    \includegraphics[width=0.6\linewidth]{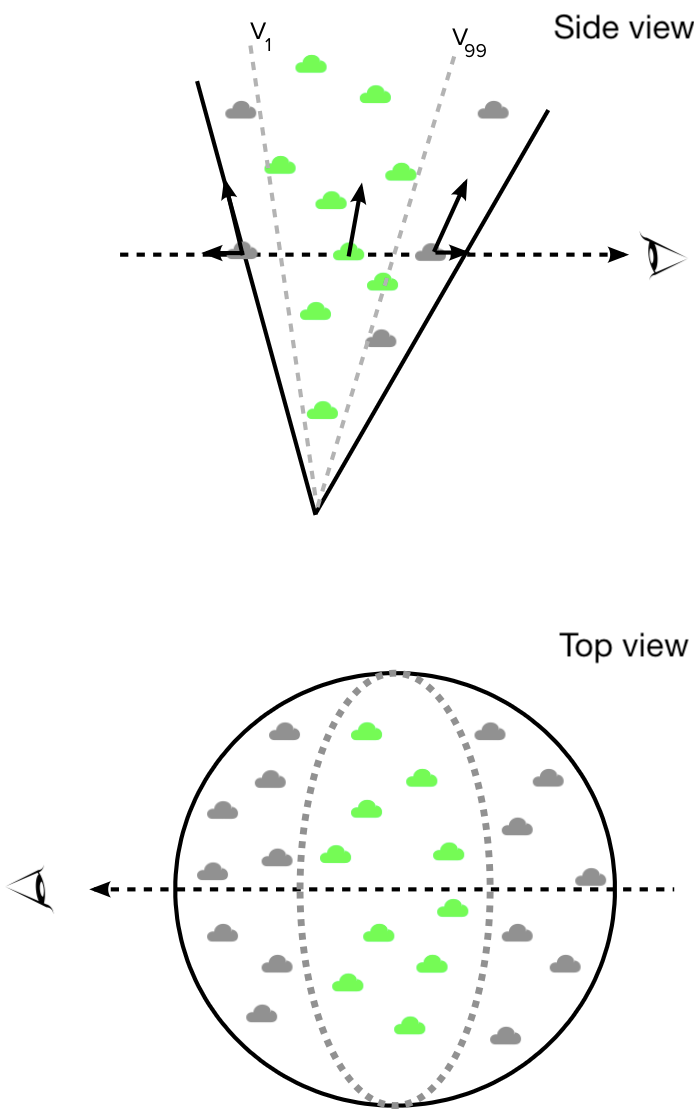}  
    \caption{Schematic explanation of the effect of an excessive radial outflow velocity with fixed model aperture and inclination. The solid black lines define the outflowing cone aperture, the dotted black lines the LOS directed towards the observer and the grey dotted lines the region containing the clouds with projected velocity within the $v_{\rm 1}$, $v_{\rm 99}$ observed boundaries. The gray clouds have LOS velocity above $v_{\rm 1}$, $v_{\rm 99}$, i.e. their weights are zero. Black arrows show the decomposition of the outflow velocity. Top panel show the view from the top of the model, bottom panel show the flattening effect as seen from the plane of the sky.}
    \label{figura_vbound}
\end{figure}

Figure \ref{figura_vbound} schematically shows the effect on model clouds distribution by creating a model with radial outflow velocity much higher than the observed one.
The solid and dotted black lines define the intrinsic outflow aperture and the LOS direction, respectively. The grey dotted lines mark the region containing clouds with projected velocity within the $\rm v_{1}$, $\rm v_{99}$ boundaries. The gray clouds, as opposite to the green ones, have projected velocities above the observed boundaries, thus the model assigns them a null weight. Top and bottom panels show the side and top view of the model, respectively.
The larger the outflow velocity, with respect to the maximum observed velocities, the more flattened on the plane of the sky is the distribution of model clouds with non-zero weight.\par
To overcome this degeneracy issue we constrained the model parameters as follows.
While following the loop over the free parameters (see scheme in Fig. \ref{model_scheme}), the algorithm discards all the combinations of model parameters which result in having percentile velocities $\rm v_{1}, v_{99}$, derived from the integrated unweighted model spectrum, different from the observed ones. This is done by defining a parameter $f$ and imposing for each model that $ \rm |v_{1}| \left( 1-f \right) < |v^{MOD}_{1}| < |v_{1}|  \left( 1+f \right)$ and $\rm |v_{99}| \left( 1-f \right) < |v^{MOD}_{\rm 99}| < |v_{99}| \left( 1+f \right)$. Here, $\rm f \in \left( 0.2, 0.05 \right)$ is a refinement parameter that progressively decreases until reaching $f = 0.05$. In this way, it is guaranteed that the difference between the observed and model percentile velocities differs of no more than $5 \%$.
This constraint then requires a suitable combination of outflow velocity, cone aperture and inclination, to reproduce the observations, besides representing an optimal method to remove the degeneracies.

\section{\MOKA \ test on simulated data}\label{mock_tests}
In this section, we present the results of the application of \MOKA \ to simulated data cubes, in order to test  model capabilities and limits.
We generated four different simulated cubes with surface brightness distribution and kinematics similar to those observed in nearby and high redshift sources. The purpose is to simulate the observed spatial and spectral resolution of new generation IFU and test the degree of reliability of the model with increasing complexity. We created each simulated cube using a spatial and spectral sampling very similar to MUSE data cubes, that is 0.2 '' / pixel and 55 km s$^{-1}$, respectively.\par
We tested different inclinations with respect to the LOS, inner and outer cone opening angles, velocity fields and PSFs, in order to simulate the outflow features observed with MUSE in our sample. Adopting different PSF sizes allows to simulate different instrument resolutions and source distances, in order to test the performance of \MOKA \ for both low- and high-z sources.
For the surface brightness emission we adopted a random distribution of 100 clumps, which is a sufficient number to produce moment maps similar to those of the MUSE sample. The clump is a 3D normal distribution of increased flux centered around a randomly selected (r, $\theta$, $\phi$) coordinate, between the minimum and maximum of the spherical coordinates interval chosen. The outflow is characterize by a constant surface brightness distribution, which is made irregular by the presence of the clumps. We want to stress that the number of simulated clumps is irrelevant for the result of the fit.
\begin{table*}
    \centering
    \begin{tabular}{c|cc|cc|cc|cc|cc|cc|cc|cc}
    \hline
    \hline
    \multicolumn{1}{c|}{} & \multicolumn{2}{c|}{$\beta$ [$^{\circ}$] } & \multicolumn{2}{c|}{$\rm v_{r}$ [km s$^{-1}$]} & \multicolumn{2}{c|}{$\rm v_{\theta}$ [km s$^{-1}$]}& \multicolumn{2}{c|}{$\rm v_{\phi}$ [km s$^{-1}$]} & \multicolumn{2}{c|}{$\theta_{\rm IN}$ [$^{\circ}$]} & \multicolumn{2}{c|}{$\theta_{\rm OUT}$ [$^{\circ}$]}  & \multicolumn{2}{c|}{P.A. [$^{\circ}$]}  & \multicolumn{2}{c}{PSF  ['']}\\
    & Sim & Fit & Sim & Fit & Sim & Fit & Sim & Fit & Sim & Fit & Sim & Fit & Sim & Fit & Sim & Fit\\
    \hline
    Model 1 &  127 & 124 & 670 & 615  & 405 & 450 & 370 & 400 & 0$^{*}$ & // & 43 & 38 & 90$^{*}$ & // & 0.5$^{*}$ & // \\
    Model 2 & 10 & 10 & 230 & 298 & -300 & -245 & 0 & 0 & 0 & 30 & 80 & 83 & 90$^{*}$ & // & 0.8$^{*}$ & //  \\
    Model 3 & 88 & 88 & 685 & 601 & 0 & 80 & 0 & -5 & 35 & 35 & 45 & 43 & 60 & 60 & 0.6$^{*}$ & //  \\ 

    \hline
    \end{tabular}
    \caption{Simulated and best fit data cube kinematical and geometrical parameters. From left to right: axis inclination with respect to the LOS ($\beta$), outflow radial velocity ($v_{\rm r}$), velocity component directed towards $\theta$ direction ($\rm v_{\theta}$) and $\phi$ direction ($\rm v_{\phi}$), inner and outer opening angles ($\theta_{\rm IN}$, $\theta_{\rm OUT}$), position angle (P.A.) and PSF in arcsec. The asterisks indicate the parameters kept fixed during the fitting procedure.}
    \label{tab.0}
\end{table*}
\noindent
For each simulated cube we inferred the best geometrical and kinematic parameters by running the fitting algorithm outlined in Sect. \ref{kin_mod_chapter}, progressively increasing the number of free parameters.\par
\begin{figure*}
\centering
\begin{subfigure}[b]{\textwidth}
   \includegraphics[width=1\linewidth]{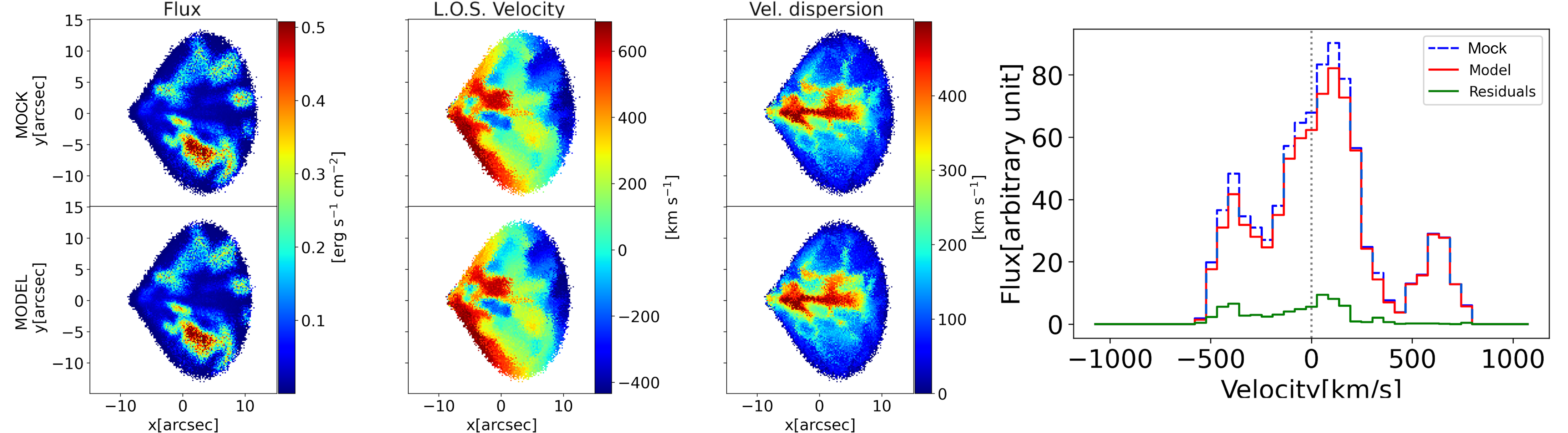}
   \label{fig:Ng1} 
\end{subfigure}

\begin{subfigure}[b]{\textwidth}
   \includegraphics[width=1\linewidth]{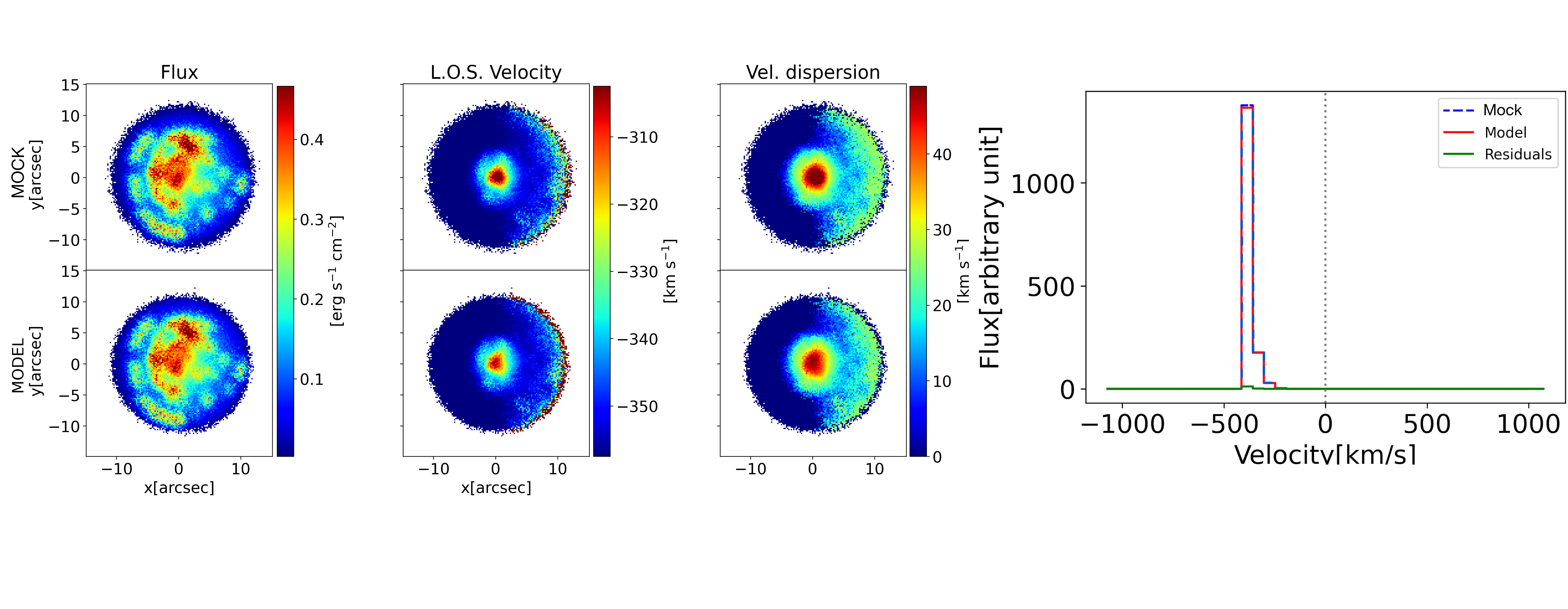}
   \label{fig:Ng2}
\end{subfigure}

\begin{subfigure}[b]{\textwidth}
   \includegraphics[width=1\linewidth]{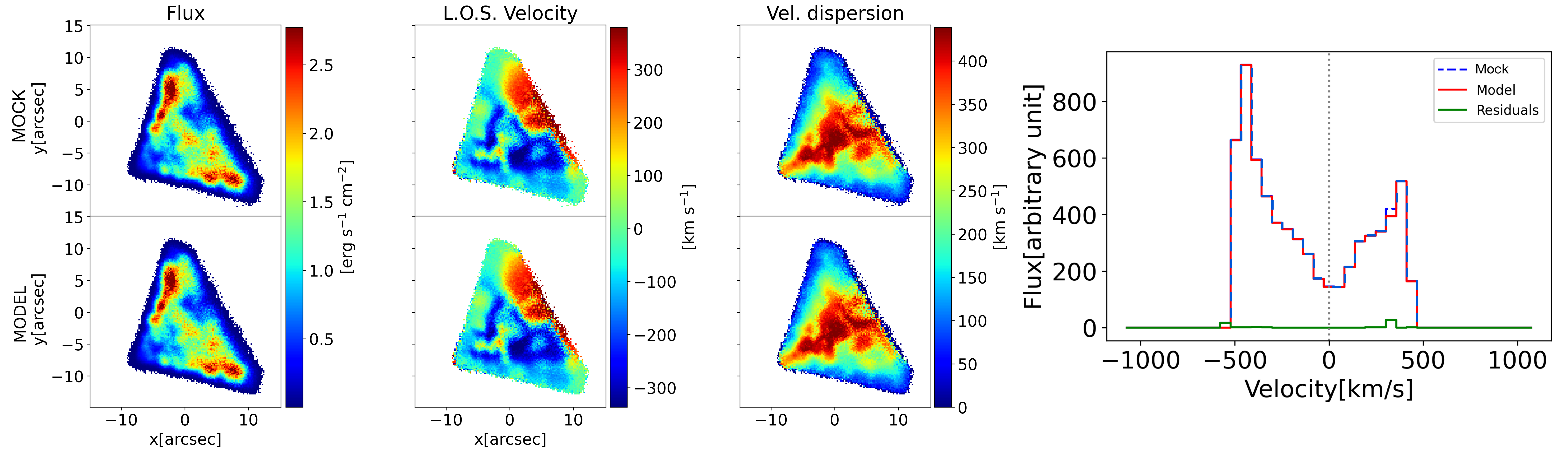}
   \label{fig:Ng3}
\end{subfigure}
\caption{From top to bottom: Model 1, Model 2 and Model 3, whose parameters are listed in Table \ref{tab.0}. Left panel: comparison of simulated cube and best fit moment maps, from left to right the integrated emission, LOS velocity and velocity dispersion maps. Right panel: comparison of simulated data (dotted blue) and best fit integrated profile (red). Residual emission is shown in green.}
\label{figura_mock}
\end{figure*}

The comparison between simulated and best fit moment maps for the first three models is shown in Fig. \ref{figura_mock}, the intrinsic and best fit parameters are reported in Table \ref{tab.0}.
Both the best fit moment maps (Fig. \ref{figura_mock}) and model parameters (Table \ref{tab.0}) are in good agreement with the simulated data. 
The first moment (velocity) maps of each simulated cube clearly show that assuming a very simple velocity field and irregular flux emission, results in very complex and irregular kinematic features, even though the intrinsic velocity field is regular.
For the first three models the fitted parameters are listed in Table \ref{tab.0}, for Model 4 instead, the parameters are reported in Table \ref{tab.out_disk}. The parameters labeled with an asterisk were kept fixed during the fit.\par
In the right panel of Fig. \ref{figura_mock} we show the comparison between simulated and best fit integrated emission line profiles, plotted in dashed blue and solid red, respectively.
The line profile residuals shown with the solid green line are to be mainly ascribed to the discrepancies in model inclination and inner/outer opening angle, which result in missing clouds at the correct 3D position, as explained in Sect. \ref{fit_subs}. \par
Model 4 was created as more similar as possible to the configuration that we expect in our MAGNUM sample. In particular, we simulated a constant radial velocity outflow, characterised by clumpy emission, superimposed on a rotating gas disk extending above the Field of View (FOV). We tested the \MOKA \ capabilities by fitting the total simulated cube emission with a simple conical model and adopting a constant radial velocity field, to test whether the model is able to derive the real kinematical and geometrical outflow parameters, despite the presence of an underlying disk. We simulated an uniform disk emission with an intensity $\sim 10^{-3}$ times smaller than the emission of the outflow clumps, and adopted a constant rotating disk velocity of 150 km s$^{-1}$, consistent with the observed stellar and gas rotating velocities in the inner kpc of galaxy disks \citep{Sofue_2015, Yoon2021, Vuckcevic2021}. The disk axis is tilted by $5^{\circ}$ with respect to the outflow axis. For Model 4 the total  emission was convolved with a spatial PSF of 0.6''.\par
\begin{figure*}
  \includegraphics[width=\textwidth]{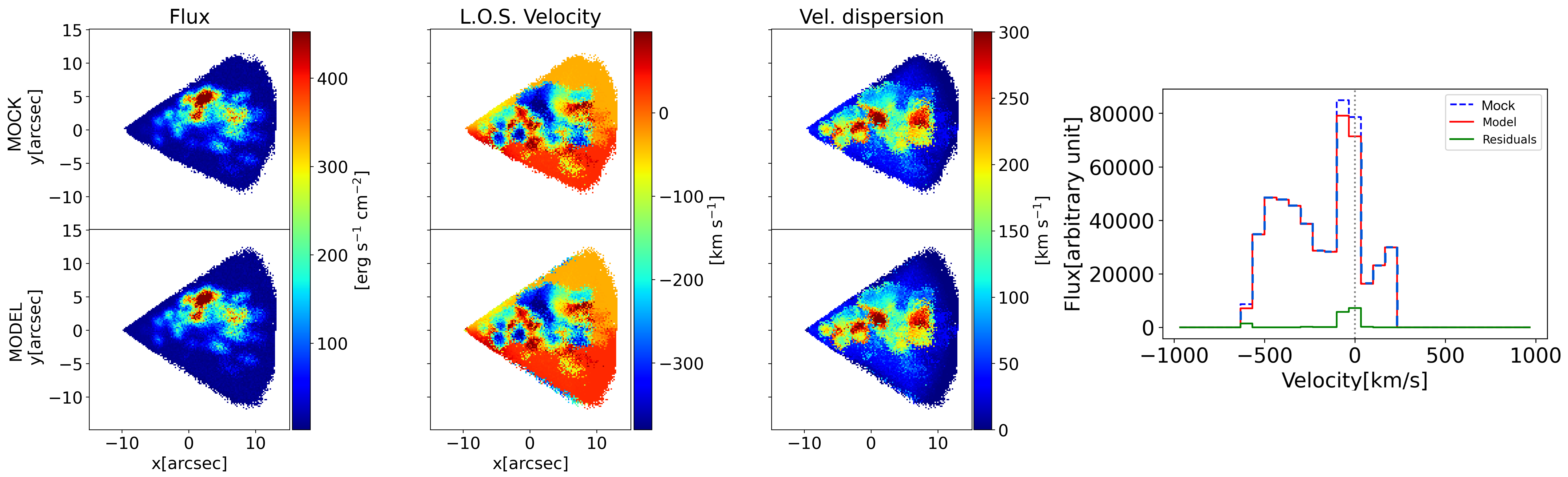}
  \caption{Left panels: Moment maps comparison of simulated data (top) and best fit model (bottom) obtained with 4 free parameters. From left to right: integrated emission, LOS velocity and velocity dispersion maps. Right panel: Integrated emission line comparison between simulated (dotted blue) and best fit model (red), the residuals are in green.
  \label{figura_out+disk}}
\end{figure*}

Figure \ref{figura_out+disk} shows the comparison between simulated and best fit model moment maps and emission line residuals, on left and right panel, respectively.
The best fit model in Fig. \ref{figura_out+disk} is obtained by fitting the outflow inclination ($\beta$), radial velocity ($\rm v_r$), inner and outer opening angle ($\rm \theta_{IN}$ and $\rm \theta_{OUT}$) (see Table \ref{tab.out_disk}).
From this test case it emerges that, in the case of combined emission of multiple kinematical components (disk and outflow), fitting only the dominant outflow component still provides accurate results. The line profile residuals in Fig. \ref{figura_out+disk} show that the best fit model is missing clouds centered at 0 km s$^{-1}$, which are due to underlying rotating disk which was not included in the model.\par
We tested models with an extremely wide range of axis inclinations, inner and outer opening angles, random clumps distribution and velocity fields, in order to account for a variety of possible configurations.
From all these tests it emerges that \MOKA\ correctly derives the kinematic and geometric parameters we used to create the simulated data cubes, with the typical PSF in low- and high-z data. In particular our model correctly derives, with unprecedented accuracy, the outflow inclination with respect to the LOS, the 3D velocity field and the outer opening angle. These represents the fundamental parameters to be measured to achieve accurate estimates of the outflow energetics. As expected, increasing the number of free parameters of the fit causes an increase of the degeneracies, with a corresponding lower accuracy even with the method outlined in Sect. \ref{degeneracies}.\par
Fitting all the outflow parameters listed in Table \ref{tab.0} or \ref{tab.out_disk}, including those labeled with the asterisk, allows for many possible configurations and unreliable best fit parameters. For example, for Model 1 and 2, we observed that keeping all the parameters free has the main consequence of providing an hollow conical best fit model, despite the intrinsic inner opening angle is exactly zero.
We observed that for full conical simulated data, fixing the inner opening angle to zero, results in an optimal estimate of the other parameters (see parameters of Model 1). Letting the inner opening angle to be a free parameter instead lead to a wrong hollow cone configuration (see Model 2). As shown for Model 3 instead, starting from hollow conical simulated data results in a correct estimate of the inner opening angle, without the need of fixing any parameter.\par
Model 4 represents the most reliable test, due to the co-presence of disk and outflow emission. For this case we decided to model only the outflow emission, exactly as we did for the MUSE sample (see Sect. \ref{magnum_test}). As shown by the best fit parameters (Table \ref{tab.out_disk}) and the moment maps comparison (Fig. \ref{figura_out+disk}), \MOKA \ represents a reliable tool to derive the outflow properties with great accuracy.

\begin{table*}
    \centering
    \begin{tabular}{c|cc|cc|cc|cc|cc|cc|cc|}
    \hline
    \hline
    \multicolumn{1}{c|}{} & \multicolumn{2}{c|}{$\beta$ [$^{\circ}$] } & \multicolumn{2}{c|}{$\rm v_{r}$ [km s$^{-1}$]} & \multicolumn{2}{c|}{$\rm v_{\theta}$ [km s$^{-1}$]} & \multicolumn{2}{c|}{$\rm v_{\phi}$ [km s$^{-1}$]} & \multicolumn{2}{c|}{$\theta_{\rm IN}$ [$^{\circ}$]} & \multicolumn{2}{c|}{$\theta_{\rm OUT}$ [$^{\circ}$]}  & \multicolumn{2}{c|}{P.A. [$^{\circ}$]}\\
    & Sim & Fit & Sim & Fit & Sim & Fit & Sim & Fit & Sim & Fit & Sim & Fit & Sim & Fit  \\
    \hline
    Model 4 &  75 & 77 & 773 & 740  & 0$^{*}$ & // & 0$^{*}$ & // & 0 & 10  & 35 & 35  & 90$^{*}$ & // \\

    \hline
    \end{tabular}
    \caption{Simulated and best fit data cube kinematical and geometrical parameters for a combination of co-spatial outflow and rotating disk. From left to right: outflow axis inclination respect to the LOS ($\beta_{\rm O}$), particles radial velocity ($v_{\rm r}$), velocity component directed towards $\theta$ direction ($\rm v_{\theta}$) and $\phi$ direction ($\rm v_{\phi}$), inner and outer opening angles ($\theta_{\rm IN}$, $\theta_{\rm OUT}$), position angle (P.A.). The simulated cube is created by convolving the emission with a PSF = 0.6''. The asterisks indicate the parameters kept fixed during the fitting procedure.}
    \label{tab.out_disk}
\end{table*}

\subsection{\MOKA\ limits}\label{moka_limits}
Based on the tests we run on simulated data, it emerged that our \MOKA\ method has limitations on the derivation of the kinematical and geometrical properties of simulated complex outflows data.
In particular, based on the results obtained for Model 4 (see Fig. \ref{figura_out+disk} and Table \ref{tab.out_disk}), which is created in order to simulate a typical data cube from the MUSE sample, it emerged that degeneracies increase with an increased number of free model parameters.
For Model 4 we tested the accuracy of our method by progressively increasing the number of free parameters and compute, voxel by voxel, the difference between the best fit model and simulated cube.
We define the accuracy as:
\begin{equation}
    A = \frac{1}{N} \sum_{i} \frac{I_{simulated, i}}{I_{fit, i}} 
    \label{accuracy}
\end{equation}
\noindent
where $I_{simulated, i}$ and $I_{fit, i}$ are the intensity of the i-th voxel of the masked simulated cube and the best fit cube, respectively. N is the total number of unmasked voxels. By definition, with the accuracy $A$ approaching unity, the corresponding model can be considered a good representation of the simulated data cube. Moreover, as a consequence of the weighting procedure (see Sect. \ref{weighted_model}), the intensity of any voxel of the weighted model cube will always be smaller, or equal, to the corresponding intensity of the simulated/observed data cube ($I_{simulated, i} \ge I_{fit, i}$).
We fitted the simulated cube with an increasing number of free parameters, starting with the radial outflow velocity ($\rm v_r$), then adding the inclination with respect to the plane of the sky ($\beta$), inner opening angle ($\theta_{\rm IN}$), outer opening angle ($\theta_{\rm OUT}$), position angle (P.A.), rotation and expansion/contraction velocity ($v_{\theta}, v_{\phi})$.
For the test case of Model 4, we observed that the accuracy level remain $> 95 \%$ fitting up to the first four outflow parameters. Adding the fifth free parameter, the accuracy drops to $\rm \sim 80 \%$. The worst scenario is obtained when fitting all the seven parameters, which results in an accuracy level of $\sim$ 60\%.\par
The test case of Model 4 is well representative of what happens fitting only the outflow conical model, despite the presence of an underlying rotating galaxy disk. 
In particular, fitting up to four outflow kinematical and geometrical parameters, results in a best fit model cube which, in each voxel, is in very good agreement with the simulated cube.
\begin{figure}
    \centering
    \includegraphics[width=\linewidth]{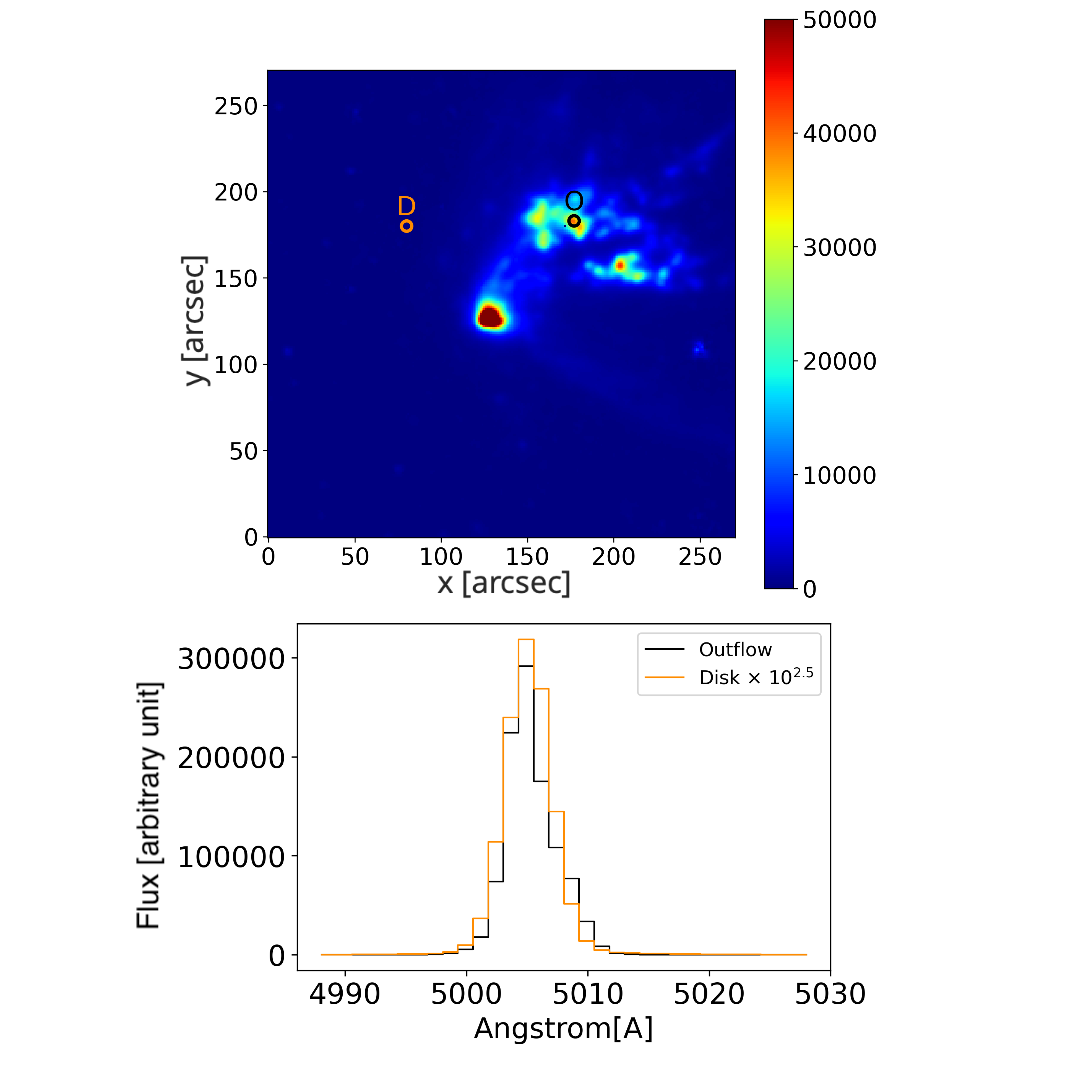}  
    \caption{Top panel: 2D [OIII]$\lambda$5007 emission map of the Circinus galaxy. Orange and black circles represents the average regions where the disk and outflow integrated emission are extracted, respectively. Bottom panel: Comparison of outflow and disk integrated spectra ($\times$ 10$^{2.5}$) extracted from the corresponding regions shown on the left panel.}
    \label{disk_out_flux_ratio}
\end{figure}
\noindent
Moreover, we observed that \MOKA \ correctly derive up to four outflow parameters, despite the presence of an underlying disk, when the observed outflow to disk flux ratio is $\geq$ 10$^{2.5}$. Figure \ref{disk_out_flux_ratio} shows the comparison of the integrated [OIII]$\lambda$5007 emission extracted from a clump of the outflow and disk, in the Circinus galaxy. The disk emission is multiplied by 10$^{2.5}$. Therefore, adopting in our simulations an average flux ratio of 10$^{2.5}$-10$^{3}$ is in agreement with the value observed in our MUSE sample. We expect that in cases with lower values of outflow to disk flux ratio, \MOKA \ would not be able to properly recover the intrinsic outflow properties. The combination of outflow and disk fitting will be addressed in future work.
\section{\MOKA \ application to MAGNUM galaxies}\label{magnum_test}
In order to show the application of \MOKA\ to real sources, we selected a sample of three nearby Seyfert-II galaxies, from the MAGNUM survey, namely NGC 4945, Circinus and NGC 7582.
In this section, we briefly introduce the main properties and gas kinematical features of this sample.

\subsection{Preliminary spectroscopic analysis}\label{preliminary_sp_analysis}
Datacubes were analysed by means of a set of custom python scripts to first subtract stellar continuum, and then fit multiple Gaussian components to the emission lines, thus finally obtaining an emission-line model cube for each emission line.
For a more detailed description of the data reduction and
the spectroscopic analysis we refer to \cite{Mingozzi2018, Marasco2020, Tozzi2021}.
\subsection{Ionized emission}
For each galaxy we extracted a sub-cube centered on the nuclear region where the ionization cone is observed. In the case of Circinus and NGC 7582 we used the [OIII]$\lambda5007$ emission line to map outflow properties; for NGC 4945 instead, we used [NII]$\lambda6584$, since [OIII] emission is highly obscured by dust in this edge-on galaxy (see e.g. Fig. A1 in \citet{Mingozzi2018}). We computed the S/N of the maps, by considering the ratio between the peak of the fitted emission line and the rms of the fit residuals, then we applied a S/N threshold of 3.
The sub-cube was extracted from the emission line model cube corresponding to the sum of all Gaussian components used to fit the line, after deconvolving the [OIII] line profile by the MUSE instrumental broadening\footnote{This is necessary since our purpose is to untangle the model results from the instrument broadening effects.}.
Once we limited the area we are interested in, we created moment maps of the ionized emission from the cube fit: the integrated line flux (moment of order 0), the flux-weighted LOS velocity (moment of order 1) and the velocity dispersion maps (moment of order 2). These maps will be used later to provide a first comparison with the final model results.
In this way, we want to demonstrate that correctly reproducing the three moment maps is a necessary but not sufficient condition to assume a model as faithful representation of the outflow features and properties; a more detailed discussion about degeneracies is carried out in Sect. \ref{degeneracies}.

\subsection{Moment maps}\label{magnum_moment_maps}
\begin{figure*}
  \includegraphics[width=\textwidth,height=15cm]{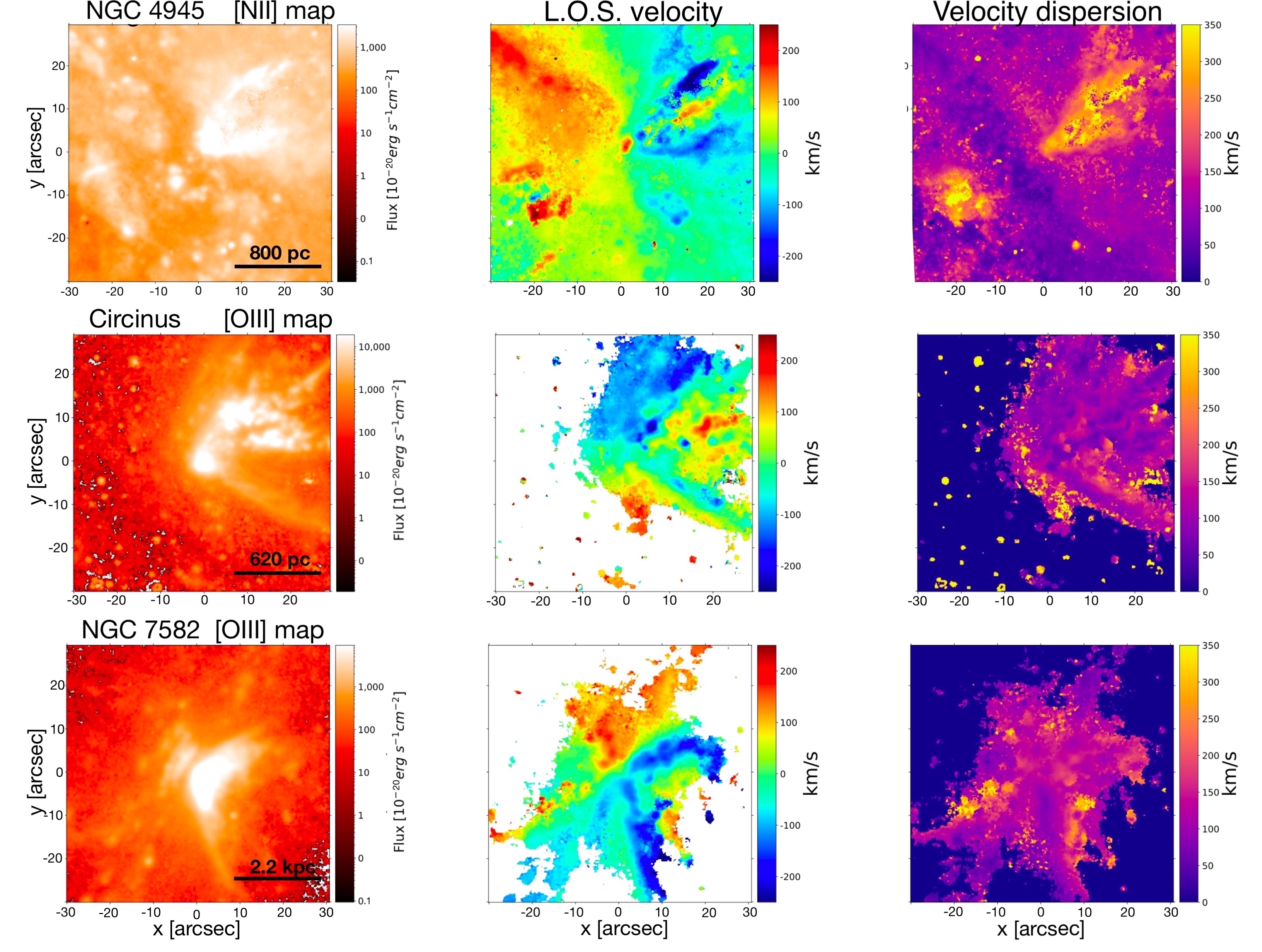}
  \caption{From top to bottom: moment maps for NGC 4945, Circinus and NGC7582. From left to right: integrated surface brightness map in unit of $10^{-20}$ erg  cm$^{−2}$ s$^{-1}$ arcsec$^{-2}$, intensity-weighted velocity and velocity dispersion. The last two rows refers to the [OIII]$\lambda 5007$ emission, the first row shows the [NII]$\lambda6584$ emission in NGC 4945. North is up. The maps are 1 spaxel-$\sigma$ spatially re-smoothed, so as to get a better visual output, and pixels with S/N < 3 are masked. The maps are obtained from the spectroscopic analysis of the modelled emission line from MUSE data \citep[e.g.][]{Marasco2020}.
  \label{figura4.3}}
\end{figure*}

In Fig. \ref{figura4.3}, we show the  moment maps for each source: starting from the left, integrated emission line, LOS velocity and velocity dispersion maps; from top to bottom: NGC 4945, Circinus and NGC 7582.
The selected galaxies were chosen to show extremely clumpy, spatially resolved ionized outflows, as highlighted by integrated emission maps in Fig. \ref{figura4.3}.
The projected distance covered by the outflow ranges from $\sim 1$ kpc in Circinus, to $\sim 3$ kpc in NGC 7582 \citep{Juneau2022}. The intensity maps show a bi-conical axis-symmetric geometry in all cases except for Circinus, where dust and gas in the galaxy disk are probably obscuring the counter cone. The receding cone in NGC 7582 and NGC 4945 is  more dust obscured than the approaching one, as  usually observed in Seyfert galaxies and AGN in general.
A common feature that we detected in all sources is a clear and steep velocity gradient across the cone, with average flux-weighted velocities from  $v < -200 $ km s$^{-1}$ at the edges, to $v > 100 $ km s$^{-1}$ at the center, along the outflow axis. Such a steep velocity gradient is commonly associated with a hollow conical geometry \citep{Venturi2017, Venturi2020}. 
The velocity dispersion of NGC 7582 and Circinus has deep minima along the galaxy plane, while is larger along the outflow extension, with sporadic peaks of $\sim 350$ km s$^{-1}$. NGC 4945 instead is characterised by a smoother and regular pattern, with the line-width increasing up to $\sim 400$ km s$^{-1}$ along the outflow axis and then decreasing towards the edges down to $\sim 250$ km s$^{-1}$, suggesting a more compact outflow geometry. 
We defined the systemic gas velocity in each source as the fitted stellar velocity, assuming co-rotation of gas and stars in the galaxy disk.

\subsection{Outflow models}\label{results_section}
\begin{figure*}
    \centering
    \begin{subfigure}{\textwidth}
        \includegraphics[width=\linewidth]{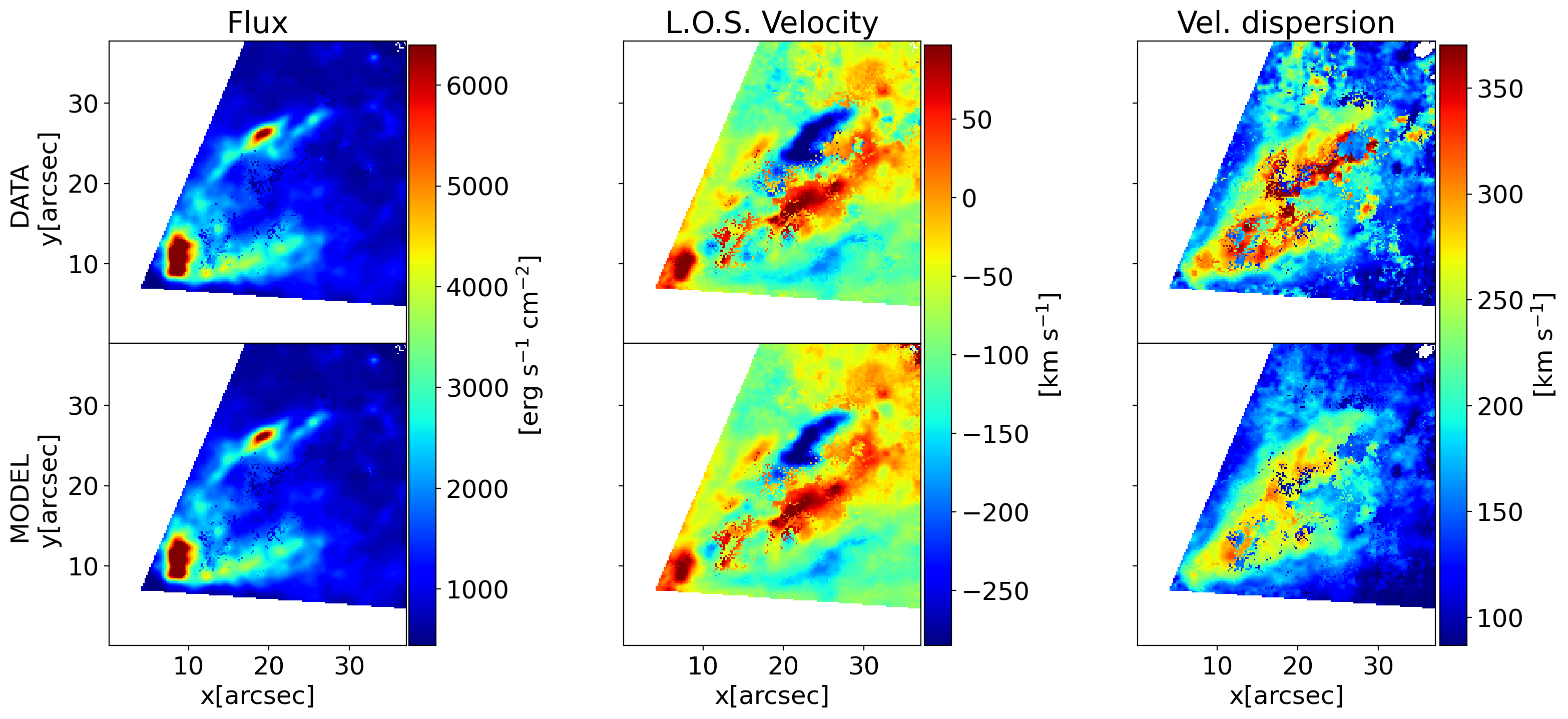}
    \end{subfigure}
    
    \medskip
    \medskip

    \begin{subfigure}{\textwidth}
        \includegraphics[width=\linewidth]{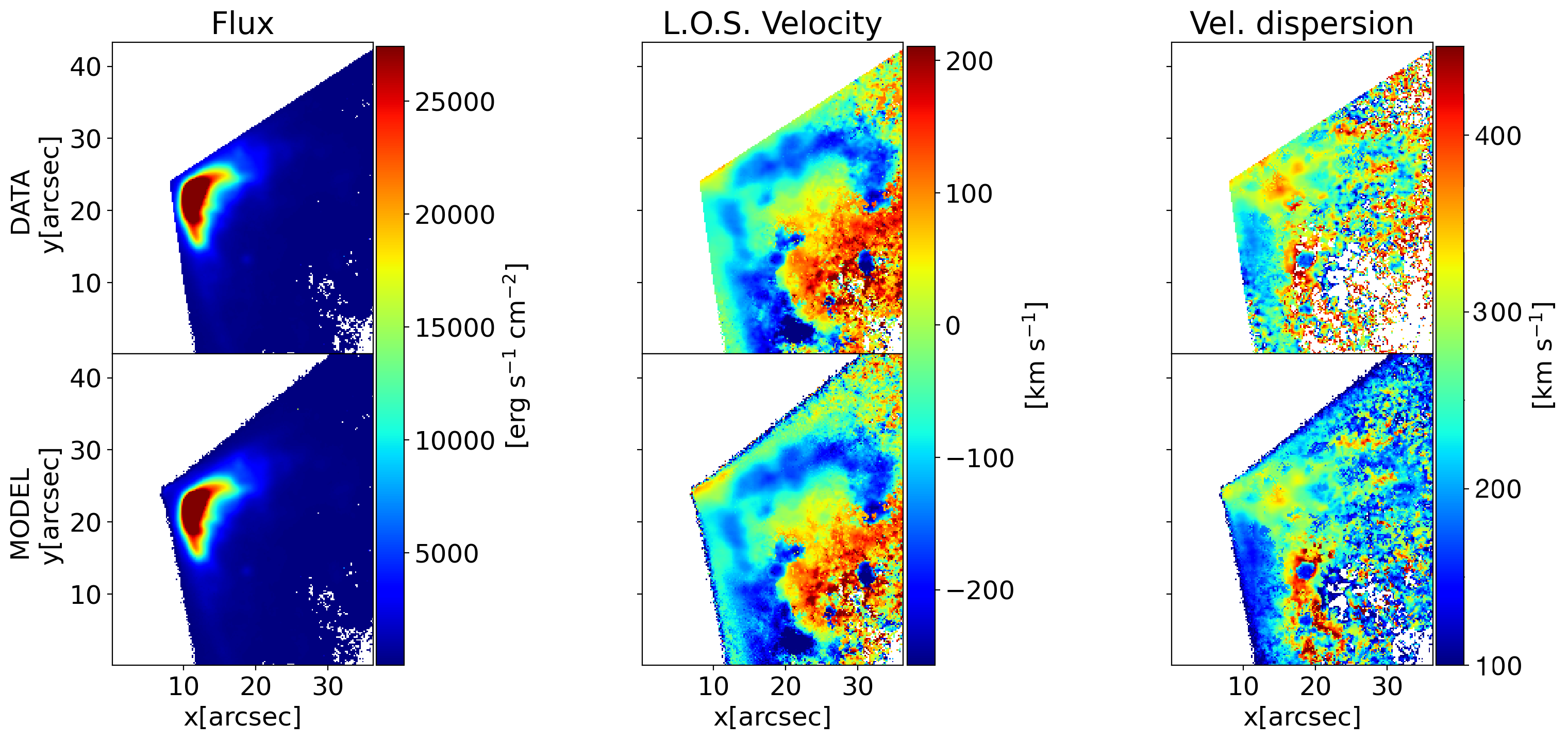}
    \end{subfigure}
    \caption{Comparison between modelled moment maps and masked observations. From left to right, panels show the integrated flux, the LOS velocity and the velocity dispersion. Conical model for NGC 4945 (a) and NGC 7582 (b). Different models for Circinus are shown in Fig. \ref{figurona}}
    \label{figura4.4}
\end{figure*}

\begin{figure*}
    \includegraphics[width=\textwidth, height=22cm]{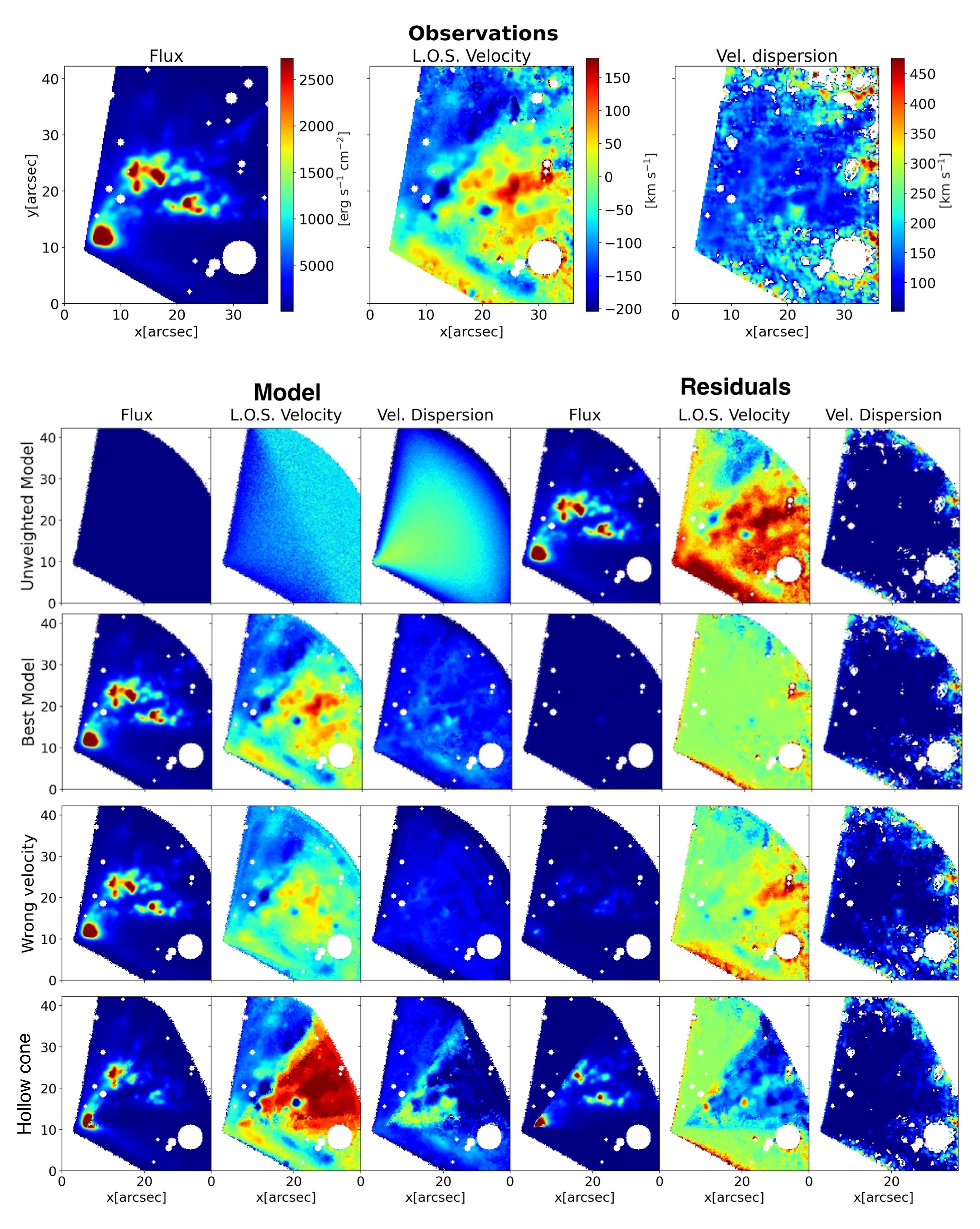}
    \caption{Different model configuration for Circinus galaxy, from left to right the ionized  [OIII]$\lambda5007$ emission line flux, the LOS velocity and velocity dispersion map. From top to bottom: masked observed data, best fit model obtained with the parameters listed in Table \ref{tab.1}, unweighted model, model with low maximum outflow radial velocity of 250 km s$^{-1}$ and hollow conical model with inner opening angle of 27 $^{\circ}$. Each model map has the corresponding residual map on the right side. The maps are color-coded according to the color-bar scale of the observed data on the top panel.
    \label{figurona}}
\end{figure*}

In this section, we present the results of our modelling applied to the three selected galaxies.
Given the best set of parameters, we first created the corresponding model and provided spatially resolved estimates of the outflow energetics, and finally compared it with the volume-averaged results obtained with the standard literature method, described later in Sect. \ref{Standard_method_energetics} \citep{CanoDiaz2012, Cresci2015b, Fiore2017, harrison2018}.
From the data we fixed both the position angle ($\gamma$, measured counter-clockwise from the north direction), the outflow conical aperture and the centre of the outflow model.
$\gamma$ coincides with the projected approaching outflow axis, the outflow aperture and centre instead, are identified by the ionized gas velocity and velocity dispersion maps. 
The outflowing cone aperture and centre are defined by the mentioned moment maps since they provide a better constraint to the projected shape and outflow starting point, rather than the clumpy integrated emission.
Then, we run the fit of the [OIII] observed emission line, as described in Sect. \ref{kin_mod_chapter}, accounting for degeneracies and minimizing the differences between modelled and observed spectra.
To estimate the outflow velocity uncertainty, we fixed all the parameters except the outflow velocity, which was varied until measuring a variation from the minimum of the $\kappa$ estimator of $10\%$ (Eq. \ref{kappa_goodness}).
Figure \ref{figura4.4} shows a comparison between the observed moment maps and those obtained with our best-fit models for NGC 4945 and NGC 7582; the observed data are masked to facilitate a comparison with the modelled maps. The same maps for Circinus are reported in first and third panels from top in Fig. \ref{figurona}.\par
The grainy emission and complex observed velocity fields of each outflow are extremely well reproduced by the model, as a consequence of weighting clouds by the observed flux.
Velocity dispersion maps highlight some discrepancies, for example in Circinus where the modeled velocity dispersion does not properly reproduce the observed clumpy structure. There might be two possible explanations for this. First, the disk contribution may not be negligible, therefore we should improve the model by considering the emission from background gas in the galaxy disk.
Second, there could be intrinsic velocity dispersion in the outflow, which we have not considered. 
As reported in Table \ref{tab.1}, we found outflows with axes close to the plane of the sky (i.e. $\beta \sim 90^{\circ}$), in all sources.
This is an obvious consequence of the fact that we selected Seyfert-II galaxies with clear evidence for an extended ionization cone, which implies that the LOS is close to perpendicular to the cone axis.\par
\medskip
\begin{table*}[h]
  \centering
  \begin{tabular}{cc|cccccc}
    \hline
    \hline
     \multicolumn{2}{c|}{Galaxy} & \multicolumn{5}{c}{Fit parameters} \\
    \hline
        ID  & $D$ &$v_{\rm 0}$ & $R_{\rm 0}^{*}$ & $\beta$ &  $v_{\rm sys}$ & $\theta^{*}$ & $\gamma^{*}$  \\
     & [Mpc] & [km s$^{-1}$] & [kpc] & [$^{\circ}$] & [km s$^{-1}$] & [$^{\circ}$] & [$^{\circ}$] \\
    \hline
    NGC 4945 & 3.8 & 1200$\pm$50 & 1.73$\pm$0.02 & 70$\pm$5  & 290$\pm$20 & 38 & 55\\
    Circinus &  4.2 & 550 $\pm$20 & 1.06$\pm$0.02 & 84$\pm$2 & -70 $\pm$10 & 55 & 65\\
    NGC 7582 & 22.7 & 630 $\pm$30 & 3.28$\pm$0.06 & 88$\pm$ 2 & -190 $\pm$ 10 &60 &110 \\
    \hline
  \end{tabular}
  \caption{Galaxies properties and best fit parameters of the ionised outflows in the sample. The asterisks indicate the parameters kept fixed during the fitting procedure. From left to right: source ID, distance, gas mean radial velocity ($v_{\rm 0}$), outflow extension in kpc ($R_{\rm 0}$), inclination with respect to the LOS ($\beta$), outflow systemic velocity with respect to the galaxy ($v_{\rm sys}$), outer semi-aperture of the conical model ($\theta$), position angle measured counter-clockwise from north ($\gamma$). Semi-apertures and position angles are fixed before the fit.}\label{tab.1}
\end{table*}
As shown by the best fit models in Fig. \ref{figura4.4} for NGC 4945 and NGC 7582 and in Fig. \ref{figurona} for Circinus, outflow features are perfectly reproduced by a full conical geometry. However, since in many studies the best result is provided by a hollow conical geometry, we tested our model with hollow conical geometries \citep[e.g.][for a statystical analysis]{Crenshaw2000a, Crenshaw2000b, Das2005, Bae2016}.
Therefore, we ran a fitting procedure considering different inner opening angles to reproduce the data.
As shown in Fig. \ref{figurona} for Circinus, this configuration is not suitable to reproduce the data; the same conclusions apply to each galaxy in our sample.
We show different model configurations, compared to the observed moment maps for Circinus (top panel). From top to bottom: the unweighted model, the best fit weighted model, an example of a model with constant radial outflow velocity of $250$ km s$^{-1}$, a hollow cone model with inner opening angle of $27.5^{\circ}$.
For each model, the residual moment maps obtained subtracting spaxel by spaxel the observed and model emission are also shown.
Both the hollow cone and the low outflow velocity configurations are clearly unable to reproduce the observed emission.
Indeed, hollow cones miss clouds at small projected velocities, while the low outflow velocity model is unable to reproduce the observed line wings (see e.g. Fig. \ref{line_prof_comparison}).
As the cone fills up, the residuals improve, reaching a minimum when the cone is completely filled. 
We conclude that a full conical geometry, with slow changes in radial velocity around a mean constant value, is well-suited to reproduce the observed outflow features in our sample.

\subsection{Outflow energetics}
In this section we investigate the ionized mass and the energetics of the outflow using two different approaches: the first approach (hereafter, referred to as "standard method") follows the prescription of \citet{CanoDiaz2012} \citep[also e.g.][]{Marasco2020, Tozzi2021}) and relies on several assumptions, since outflow physical properties are usually unknown. The second (hereafter '\MOKA \ based method') is based on our novel 3D geometry and kinematic modelling.
\subsubsection{Standard method}\label{Standard_method_energetics}
We calculate the mass outflow rate through the surface of a spherical sector with radius $r$ defined by the cone aperture, using the outflow velocity inferred from the [OIII] emission line. If the outflow geometry and inclination with respect to the LOS are unknown, as done for example in \citet{Marasco2020} and \citet{Tozzi2021}, we can define the outflow velocity as: 
\begin{equation}
     \rm v_{out} = max \left( |v^{max}_{10} - v_{sys}|, |v^{max}_{90}-v_{sys}| \right)   ,   \label{v_out_marasco} 
\end{equation}
\noindent
where $\rm v^{max}_{\rm 10}$ and $\rm v^{max}_{\rm 90}$ are the maximum percentiles velocities corresponding to the $10\%$ and $90\%$ of the flux of the outflow component profile, respectively, and $\rm v_{sys}$ is the systemic velocity of the galaxy, set to $0$ km s$^{-1}$.
Equation \ref{v_out_marasco} assumes that the intrinsic outflow velocity can be determined by the wing of the line profile, which provides the maximum observed velocity, possibly correcting for projection effects and dust absorption \citep[e.g.][]{Cresci2015b}.
The [OIII] line luminosity ($L_{\rm [OIII]}$, that we measure in each spaxel from the observations) can be written from the theoretical point of view as:
\begin{equation}
    \rm L_{[OIII]} = \int_{V} n_e n\left( O^{2+} \right)\epsilon_{[OIII]}  f  dV, \label{luminosityoiii} \,
\end{equation} 
\noindent
where $V$ is the volume occupied by the outflowing ionized gas, $\rm f$ the filling factor of the [OIII] emitting clouds in the outflow, $n(O^{2+})$ and $n_e$ are the volume densities of the $O^{2+}$ ions and of
electrons, respectively, and $\epsilon_{\rm [OIII]}$ the [OIII]$\lambda5007$ emissivity, having a weak dependence on the temperature ($ \propto T^{ 0.1} $) at the typical temperature of the NLR ($\sim 10^4 $K). As done in \cite{CanoDiaz2012}, here we assume that most of the oxygen within the outflowing ionized gas is in the $\rm O^{2+}$ form, and neglect the contribute to the mass from species heavier than helium. Finally, the outflow mass can be computed as: 
\begin{equation}
    \rm M_{out} = 5.33 \times 10^{7}  \frac{C \ L_{44} \left( [OIII] \right)}{<n_{e, 3}> 10^{[O / H]}} M_{\odot} \label{m_out_final_canodiaz} \,
\end{equation}\\
\noindent
where $\rm L_{ 44} \left( [OIII] \right)$ is the luminosity of the total [OIII]$\lambda5007$ emission line profile in units of 10$^{44}$ erg s$^{-1}$, $\rm <n_{\rm e, 3}>$ is the average electron density in the ionized gas clouds in units of 10$^{3}$ cm$^{-3}$ (i.e. $\rm <n_{\rm e, 3}> = \int_V n_e \ f \ dV / \int_V  \ f \ dV $) and [O/H] represents the oxygen abundance in solar units. $\rm C=<n_{\rm e, 3}>^2/<n_{\rm e, 3}^2>$ is a “condensation factor” (for more detail on the derivation of the warm mass traced by the [OIII]$\lambda5007$ line in every pixel (see \citealt{Rupke2005, Genzel2011, Carniani2015, Cresci2017, Veilleux2020}). We can assume C = 1 under the simplifying hypothesis that all ionized gas clouds in each resolution element (a MUSE spaxel in our case) have the same density. Also, under these assumptions, the mass of the outflowing ionized gas is independent from the filling factor of the emitting clouds.
Finally we are able to infer the average mass outflow rate in the volume, as follows:
\begin{equation}
    \rm {\dot M}_{\rm out} = 164 \ \frac{ L_{\rm 44}([OIII])\ v_3}{<n_{\rm e, 3}> \ 10^{[O/H]} \ R_{\rm kpc}} \ M_{\odot} yr^{-1} \label{mass_out_rate_canodiazfinale} \,
\end{equation}
\noindent
where $v_3$ is the outflow velocity in units of $1000 $ km s$^{-1}$, and $R_{\rm kpc}$ is the conical outflow radius, in units of kpc. The outflow rate is independent of both the opening angle $\Omega$ of the outflow and the filling factor $f$ of the emitting clouds (under the assumption of clouds with the same density).
The kinetic energy ($E_{\rm kin}$), kinetic luminosity ($L_{\rm kin}$) and momentum rate (${\dot p}_{\rm out}$), of the ionised outflow, are then given by:
\begin{equation}
    \rm E_{ kin} = 9.94 \times 10^{42} \ \Big(\frac{ M_{\rm out}}{M_{\odot}}\Big) \ \Big(\frac{v_{\rm out}}{km s^{-1}} \Big)^2 \ erg, \label{ekin} \,
\end{equation}
\begin{equation}
    \rm L_{\rm kin} = 3.16 \times 10^{35} \ \Big(\frac{\dot M_{\rm out}}{M_{\odot} yr^{-1}}\Big) \ \Big(\frac{v_{\rm out}}{km s^{-1}}\Big)^2 \ erg \ s^{-1}, \label{Lkin} \,
\end{equation} 
\begin{equation}
    \rm \dot p_{ out} = 6.32 \times 10^{30} \ \Big(\frac{ \dot M_{\rm out}}{M_{\odot}yr^{-1}}\Big) \ \Big(\frac{v_{\rm out}}{km s^{-1}}\Big) \ dyne. \label{pout} \,
\end{equation}
\noindent
Eqs. \ref{m_out_final_canodiaz} - \ref{pout} require the knowledge of different physical properties of the outflow, but only a few of those are usually measured, while others have to be assumed. The quantities usually assumed are the oxygen abundance, which is usually fixed to the solar abundance, and the electron density, which can be estimated from the [SII]$\lambda\lambda6717,6731$ doublet or needs to be fixed to typical values of AGN at similar redshift of the sample.
If the S/N for the [SII] doublet is high enough and the two lines are spectrally resolved, $n_e$ can be directly estimated from the flux ratio for the lines in the doublet \citep[e.g.][]{Osterbrock2006, Kewley2019}, as follows:
\begin{equation}
    n_e = \frac{627.1 R - 909.2}{0.4315 R} \label{electrondensity} \,
\end{equation}
\noindent
where R is the flux ratio of the total emission line profile of the sulfur doublet $f([SII]\lambda6717)/f([SII]\lambda6731)$.
Assuming a constant electron density can have a huge impact on the outflow energetic \citep[e.g.][]{Kakkad2018, Davies2020a}, nevertheless this is necessary in those cases where an estimate from the data is not possible.
\begin{table*}
\centering
    \begin{tabular}{c| c c c c c c } 
    \hline
    \hline
    ID &  $\rm M_{\rm out}$ & $\rm E_{\rm KIN}$ & \rm $\rm L_{\rm KIN}$ & $\rm {\dot p}_{\rm out}$  & $\rm n_e$ & $\rm logL_{\rm bol}$ \\
       & $10^{3} M_{\odot}$ & $10^{3} M_{\odot}$ $10^{52}$ erg & $10^{40}$ erg $^{-1}$ & $10^{33}$ dyne & cm$^{-3}$ & erg s$^{-1}$ \\
    \hline
    NGC 4945 &  2.1 $\pm$ 1.2 & 3.8 $\pm$ 1.9  & 19 $\pm$ 10  & 3.1 $\pm$ 1.5  &  110 & 42.3      \\
    Circinus & 3.3 $\pm$ 1.9 & 1.2 $\pm$ 0.6  & 3.6 $\pm$ 1.8  & 1.3 $\pm$ 0.6  & 300 & 42.9 \\
    NGC 7282 &  60 $\pm$ 36 & 29 $\pm$ 15  & 27 $\pm$ 13  & 8.6 $\pm$ 4.0 & 260 & 44.1 \\
    \hline
    \end{tabular}
    \caption{Energetic properties of the outflow in the sample. From left to the right: Total kinetic energy, total kinetic luminosity, total momentum rate, electron densities and AGN bolometric luminosity in log scale. The uncertainties are computed via error propagation from Eqs. \ref{ekin}, \ref{Lkin}, \ref{pout}.} 
    \label{tab.3}
\end{table*} 

The outflow energetic properties, calculated using Eqs. \ref{m_out_final_canodiaz}-\ref{pout}, are reported in Table \ref{tab.3}, except for the ionized mass outflow rate, which will be discussed later. Their uncertainties were obtained with error propagation, for the electron density we assumed a systematic uncertainty of $50\%$ \citep[][]{Tozzi2021}. 
For NGC 7582 and NGC 4945 we determined the electron density from [SII] doublet with Eq. \ref{electrondensity}.
For Circinus, instead we averaged the value of the spatially resolved study by \citet{Mingozzi2018}, obtained from the high velocity parts of the [SII] doublet lines.
We found mass outflow rates of $1.9 \times 10^{-3} M_{\odot}$yr$^{-1}$, $0.2 \times 10^{-2} M_{\odot}$yr$^{-1}$ and $1.4 \times 10^{-2} M_{\odot}$yr$^{-1}$ for NGC 4945, Circinus and NGC 7582, respectively. 
NGC 4945 is also characterized by a molecular outflow traced by ALMA CO J = 3-2 emission, co-spatial to the ionized one, with an estimated mass outflow rate of $\sim 20 M_{\odot}$yr$^{-1}$, as reported by \citet{Bolatto2021} (Carniani et al. in prep.).
The inferred ${\dot M}_{\rm out}$ for Circinus, computed over a distance of $\sim 1$ kpc from the AGN, is smaller by a factor of $\sim 100$ compared to the estimate of \citet{Fonseca2021}. They found an outflow rate for the blue and red components of $0.12 \ M_{\odot}$yr$^{-1}$ and $0.1 \ M_{\odot}$yr$^{-1}$, respectively, by selecting the spaxels with detected high-ionisation outflow traced by [Fe vii]$\lambda6089$, and considering a maximum distance from the AGN of $700$ pc. 
The ${\dot M}_{\rm out}$ obtained for NGC 7582 is consistent with the high-ionization counter-part of $0.7 \times 10^{-2} M_{\odot}$yr$^{-1}$, inferred by \citet{Davies2020a} using VLT/Xshooter spectra and covering the inner $300$ pc. Both the outflow extension and maximum velocity of $364$ km s$^{-1}$ are much lower compared to our work, in which we estimated the outflow rate considering an outflow extension of $\sim 3.2$ kpc and a radial velocity of $\sim 630$ km s$^{-1}$.
\subsubsection{Wind energetic uncertainties}\label{wind_assumptions}
Literature estimates of the wind parameters rely on different assumptions for each gas phase.
The electron density in the outflowing gas, the temperature and geometry are among the parameters that most affect the uncertainties of the outflow energetics. As explained in Sect. \ref{Standard_method_energetics}, when it comes to estimate the ionized wind mass and energetics, many basic assumptions can affect the final parameters, leading to systematic
uncertainties of a factor of $\sim$ 10. Therefore, even adopting the standard recipe outlined in Sect. \ref{Standard_method_energetics}, it is not unexpected to observe up to three orders of magnitude of scatter for outflow energetics in different studies\citep[e.g.][]{Fiore2017, Cicone2018, harrison2018}.
In this context, our \MOKA\ method represents an innovative approach to determine the outflow physical properties with great accuracy, since it is not based on strong observational or physical assumptions.
\subsubsection{\MOKA \ based method}\label{new_method}
\begin{figure*}[h!]
    \centering
    \includegraphics[width=0.85\textwidth]{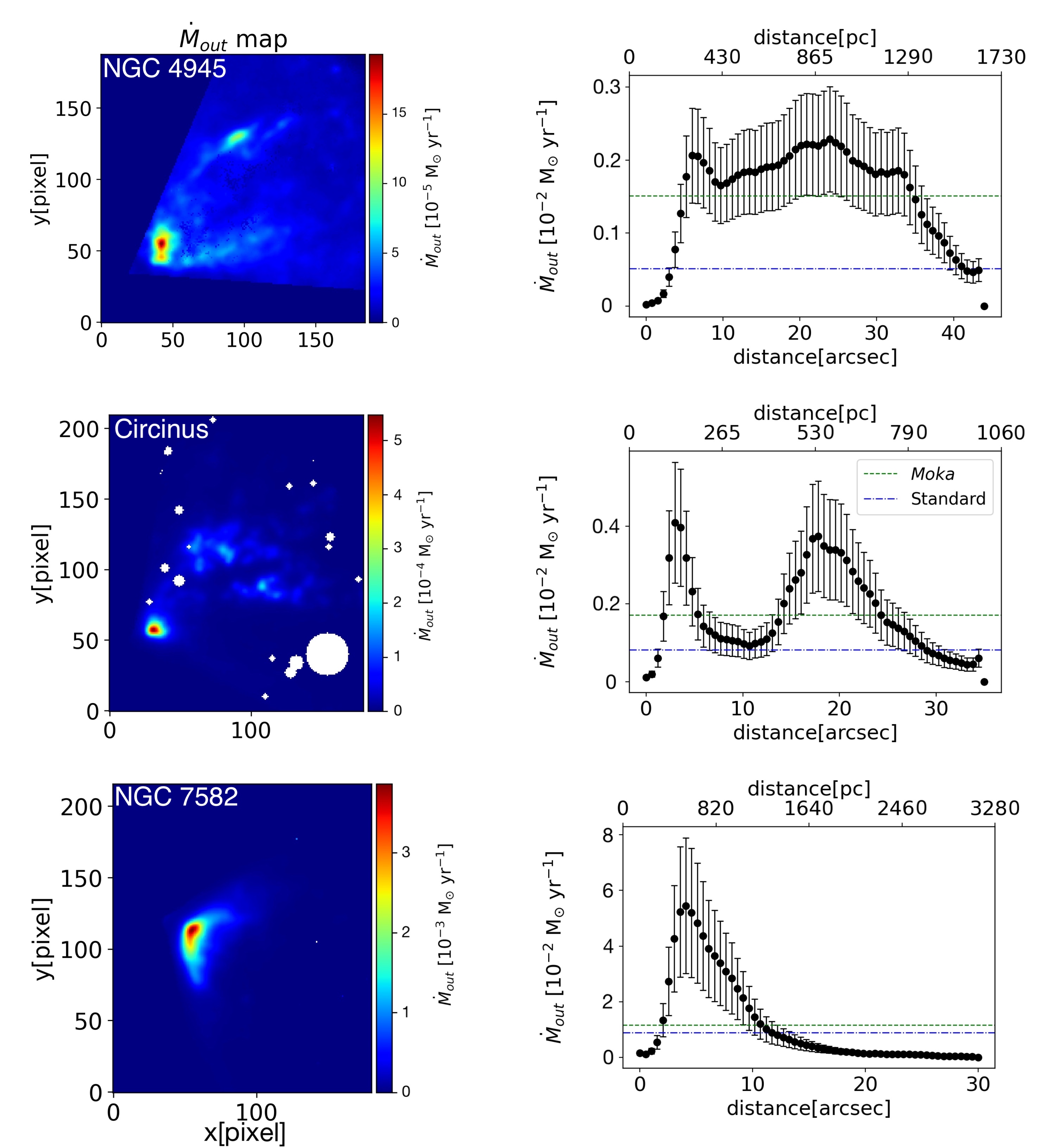}
    \caption{Left panels: 2D maps of the mass outflow rate. Right panels: Mass outflow rate radial profile. Dashed green and blue lines represent the volume average mass outflow rate obtained with the \MOKA \ based method and the standard method, respectively. From top to bottom: NGC 4945, Circinus and NGC 7582.
    \label{figura4.6}}
\end{figure*}

Employing the tomographic outflow reconstruction, we have a 3D distribution of ionized gas clouds within the outflow. This allowed us to provide a spatially resolved estimate of the outflow energetics.
To compute the amount of ionized mass in each voxel we used Eq. \ref{m_out_final_canodiaz}, converting the flux density emitted by each cloud in the same bin to ionized mass in solar masses.
Assuming the flux density within each spaxel to be constant over time and the outflow to subtend a solid angle $d\Omega$, we use the continuity equation in spherical coordinates and express the ionized mass outflow rate via the following equation:
\begin{equation}
     \rm \dot M_{\rm out} =  d\Omega\, r^{2} \rho v \label{M_dot_vero}
\end{equation}
\noindent
where $\rho$ is the mass density in each 3D volume element and $v$ the outflow velocity, assumed to be constant within each spatial element.
The mass density in each voxel is $\rho = dM_{\rm ion} / d\Omega r^{2} dr$, with $dr = 0.2''$ the MUSE spatial resolution and $dM_{\rm ion}$ the ionized mass in the same spaxel.
Therefore Eq. \ref{M_dot_vero} becomes:
\begin{equation}
    \rm \dot M_{out} = \frac{dM_{\rm ion}v}{dr} \label{M_dot_vero_finale}
\end{equation}
\noindent
being ${\dot M}_{\rm out}$ the average mass outflow rate within each spatial element of width $dr$, at a certain distance from the AGN. To estimate the mass outflow rate profile, we computed the amount of ionized gas ($dM_{\rm ion}$) in a shell of fixed width ($dr$) and assumed the radial velocity to be constant within the shells. 
2D maps of the mass outflow rate, and their respective radial profiles, for NGC 4945, Circinus and NGC 7582 are shown in Fig. \ref{figura4.6} in left and right panels, respectively.
To calculate the uncertainties on ${\dot M}_{\rm out}$ in each shell, we propagated the errors, considering the inferred
error on $v$ from our modelling (see Sect. \ref{results_section}) and assuming an uncertainty of 0.6'' on $dr$.\par
We found three different radial profiles which can be linked to the past AGN accretion histories and variation of AGN luminosity, possibly indicating large time variations in ionizing luminosity that result in the peaks of the outflow rate \citep{Keel2012, Gagne2014, Keel2015, Keel2017, Finlez2022}. However, this could also be an ionization effect which is not taken into account when assuming a constant luminosity-mass conversion factor, as we have done. 
A proper determination of the ionized gas mass from the emission line luminosity requires the use of photo-ionization models, which we will address in future work.
This can also be observed by inspecting the outflow clumpiness distribution. In NGC 4945 and Circinus we observed two shells with increased flux density. In NGC 7582 instead, we observed one peak of emission slowly decreasing with the distance from the source, up to the maximum model extension.
From the mass outflow rate profile, we can estimate the dynamical timescale of the ionized outflow $t_{\rm dyn} = d_e / v_e$, where $d_e$ and $v_e$ are the maximum outflow extension and intrinsic velocity inferred with \MOKA , respectively. We obtained  $1.4 \times 10^{6}$ yr,  $1.8 \times 10^{6}$ yr and $4.9 \times 10^{6}$ yr, for NGC 4945, Circinus and NGC 7582 respectively, consistently with theoretical predictions and simulations \citep{Nardini2018, Richings2018}.
\subsubsection{Methods comparison}\label{method_comparison}
In addition to the spatially resolved mass outflow rate estimate, \MOKA \ can provide a volume-averaged estimate of $\rm \dot M_{out}$. This is done using Eq. \ref{M_dot_vero_finale} by considering the total amount of ionized gas mass, inferred by the emission of each model clouds, the maximum model distance from the AGN and the intrinsic outflow velocity.
\begin{table*}[h!]
  \centering
  \begin{tabular}{c|ccc|ccc}
    \hline
    \hline
    \multicolumn{1}{c|}{ } & \multicolumn{3}{c|}{Standard method} & \multicolumn{3}{c}{\MOKA } \\
    \hline
        ID &  $\rm {R^{std}}_{\rm max}$ & ${\rm v^{\rm std}}_{\rm max}$ & $ {\rm {\dot M}^{\rm std}}_{\rm out}$ & ${\rm R^{moka}}_{\rm max}$ & ${\rm v^{\rm moka}}_{\rm max}$ & ${\rm {\dot M}^{\rm moka}}_{\rm out}$  \\
    & kpc & km s$^{-1}$ & $\rm 10^{-2} M_{\odot}$yr$^{-1}$& kpc & km s$^{-1}$ &  $\rm 10^{-2} M_{\odot}$yr$^{-1}$\\
    \hline
    NGC 4945 & 1.63 & 410 & 0.06$\pm$0.04 & 1.73 & 1200 & 0.14$\pm$0.09  \\
   
    Circinus & 1.05 & 270 & 0.08 $\pm$0.5 & 1.06 & 550  & 0.17 $\pm$ 0.10  \\
    
    NGC 7282 & 3.27 & 390 & 0.9$\pm$0.6 & 3.28 & 630 & 1.2$\pm$0.7   \\
    \hline
  \end{tabular}
  \caption{Comparison between standard and \MOKA \ based method results of the ionized outflow maximum extension, radial velocity and volume-average mass outflow rate in the sample.}\label{tab.4}
\end{table*}
As explained in Sect. \ref{weighted_model}, each cloud has an assigned weight which is derived from the observed emission in the corresponding voxel. Therefore, we convert each cloud weight to [OIII] luminosity and then derive the ionized gas mass.
In Table \ref{tab.4} we show the parameters and the volume-average mass outflow rate, estimated via the standard method (\ref{Standard_method_energetics}) and our \MOKA \ model (\ref{new_method}).
The estimates provided by the two methods are compatible within the errors reported in Table \ref{tab.4}. Moreover, the volume-average mass outflow rates have the same order of magnitude of the radial profile, as shown in the right panels in Fig. \ref{figura4.6}. The standard method underestimates the average $\rm \dot M_{out}$ of a factor of 2, compared to our method, based on \MOKA. This is due to the fact that, both the velocity (see Eq. \ref{v_out_marasco}) and the projected maximum radius extension adopted by the standard method are smaller compared to the values provided by our model. Both methods are affected by systematic uncertainties of the order of $50 \%$, as result of the assumed uncertainties on the electron density. As discussed above, to refine the energetics estimate we plan to combine our \MOKA \ model with a photo-ionization model, to properly constrain the outflowing gas mass and further reduce the uncertainties. 

\section{Conclusions}\label{conclusions_section}
We have developed a new 3D kinematical model which allows us to constrain with unprecedented accuracy the kinematics and geometry of clumpy AGN galactic winds, reproducing the emission line features observed in all spaxels.  We applied the model to three nearby Seyfert-II galaxies selected from the MAGNUM survey \citep{Cresci2015b, Venturi2018, Mingozzi2018, Venturi2021}, featuring a clear (bi)conical ionized outflow extending over kpc scales, and showed that the observed complexity in the kinematical maps is well reproduced by a simple radial outflow model with constant velocity and a clumpy ionized gas clouds distribution.
The main features and results of our model are summarized below:

\begin{enumerate}
    \item In Sect. \ref{kin_mod_chapter}, we build a multi-cloud model to take into account the intrinsic clumpy nature of ionised outflows. We weight each of our model clouds based on the observed line emission and velocity, and constrain the input parameters through a fit by comparing the model and observed line profile spaxel by spaxel.
    \item The model takes into account the spatial and spectral resolution of the data, by creating a model cube with identical 3D extensions, spatial and spectral resolutions of observed data cubes. \MOKA \ also accounts for the convolution of model clouds surface brightness with the PSF measured from observed data.
    \item One of the main achievements of this modelling technique is to disentangle the observed features from projection effects, obtaining the real 3D distribution of the gas clouds in a tomographic way.
    \item In this work, the free outflow kinematical and geometrical parameters that the modelling retrieves are the outflow inclination with respect to the plane of the sky, the gas cloud radial velocity and the systemic velocity with respect to the host galaxy (see Sect. \ref{results_section}).
    \item In Sect. \ref{mock_tests}, we tested the capabilities of our new method by applying it to four simulated cubes with moment maps similar to the ones observed in the MAGNUM sample. The model manages to reproduce the observed features and provide the correct outflow parameters, with an accuracy > 95 \%, fitting up to four parameters.
    \item In Sect. \ref{new_method}, we managed to measure with great accuracy and without strong assumptions the outflow physical properties (i.e. mass outflow rate, kinetic luminosity, kinetic energy and momentum rate).
\end{enumerate}
With our model we provide a new reliable method to constrain the wind energetics, the powering mechanism in AGN and a different approach to infer the impact of outflows on the ISM and galaxy evolution, expanding the knowledge on AGN feedback mechanism.
In \cite{cresci2023} we present the first application of \MOKA \ to a high-z QSO, observed with JWST/NIRSpec and characterized by a collimated high velocity outflow piercing an expanding bubble of ionized gas.\footnote{For the outflow 3D reconstruction see our \MOKA \ YouTube channel: \url{https://www.youtube.com/channel/UCQ12ob3CuraQqedCNCGHUsA}}
In future work we will apply our model to the other galaxies in the MAGNUM survey, as well as to other high-redshift quasars \citep[e.g. those in][]{Carniani2015}.
We will also consider more complex models where a disk or other kinematical components contribute to the observed data cubes.
\begin{acknowledgements}
       GC, AM, GT, FM, FB and GV acknowledge the support of the INAF Large Grant 2022 "The metal circle: a new sharp view of the baryon cycle up to Cosmic Dawn with the latest generation IFU facilities". GC, AM acknowledge support from PRIN MIUR project “Black Hole winds and the Baryon Life Cycle of Galaxies: the stone-guest at the galaxy evolution supper”, contract \# 2017PH3WAT. EDT was supported by the European Research Council (ERC) under grant agreement no. 101040751. S.C acknowledges funding from the European Union (ERC, WINGS,101040227). G.V. acknowledges support from ANID program FONDECYT Postdoctorado 3200802. 
       
\end{acknowledgements}

\bibliographystyle{aa}
\bibliography{biblio} 

\begin{thebibliography}{74}
\expandafter\ifx\csname natexlab\endcsname\relax\def\natexlab#1{#1}\fi

\bibitem[{Bae \& Woo(2016)}]{Bae2016}
Bae, H. \& Woo, J. 2016, \href{https://doi.org/10.48550/arXiv.1606.05348}{ApJ},
  828, 185

\bibitem[{Bae \& Woo(2018)}]{Bae2018}
Bae, H.-J. \& Woo, J.-H. 2018,
  \href{https://doi.org/10.3847%2F1538-4357%2Faaa42d}{ApJ}, 853, 185

\bibitem[{Bolatto {et~al.}(2021)Bolatto, Leroy, Levy, Meier, Mills, Thompson,
  Emig, Veilleux, Ott, Gorski, Walter, Lopez, \& Lenki{\'{c} }}]{Bolatto2021}
Bolatto, A.~D., Leroy, A.~K., Levy, R.~C., {et~al.} 2021,
  \href{https://doi.org/10.3847/1538-4357/ac2c08}{ApJ}, 923, 83

\bibitem[{Boroson(2005)}]{Boroson2005}
Boroson, T. 2005, \href{https://doi.org/10.1086%2F431722}{ApJ}, 130, 381

\bibitem[{Bouch{\'{e} } {et~al.}(2015)Bouch{\'{e} }, Carfantan, Schroetter,
  Michel-Dansac, \& Contini}]{boiche2015}
Bouch{\'{e} }, N., Carfantan, H., Schroetter, I., Michel-Dansac, L., \&
  Contini, T. 2015,
  \href{https://doi.org/10.1088%2F0004-6256%2F150%2F3%2F92}{AJ}, 150, 92

\bibitem[{Cano-Díaz {et~al.}(2012)Cano-Díaz, Maiolino, Marconi, Netzer,
  Shemmer, \& Cresci}]{CanoDiaz2012}
Cano-Díaz, M., Maiolino, R., Marconi, A., {et~al.} 2012,
  \href{https://doi.org/10.1051/0004-6361/201118358}{A\&A}, 537, L8

\bibitem[{Carniani {et~al.}(2015)Carniani, Marconi, Maiolino, Balmaverde,
  Brusa, Cano-D{\'{\i} }az, Cicone, Comastri, Cresci, Fiore, Feruglio, Franca,
  Mainieri, Mannucci, Nagao, Netzer, Piconcelli, Risaliti, Schneider, \&
  Shemmer}]{Carniani2015}
Carniani, S., Marconi, A., Maiolino, R., {et~al.} 2015,
  \href{https://doi.org/10.1051/0004-6361/201526557}{A\&A}, 580, A102

\bibitem[{Carniani {et~al.}(2016)Carniani, Marconi, Maiolino, Balmaverde,
  Brusa, Cano-D{\'{\i} }az, Cicone, Comastri, Cresci, Fiore, Feruglio, Franca,
  Mainieri, Mannucci, Nagao, Netzer, Piconcelli, Risaliti, Schneider, \&
  Shemmer}]{Carniani2016}
Carniani, S., Marconi, A., Maiolino, R., {et~al.} 2016,
  \href{https://doi.org/10.1051/0004-6361/201528037}{A\&A}, 591, A28

\bibitem[{Cicone {et~al.}(2018)Cicone, Brusa, Almeida, Cresci, Husemann, \&
  Mainieri}]{Cicone2018}
Cicone, C., Brusa, M., Almeida, C.~R., {et~al.} 2018,
  \href{https://doi.org/10.1038%2Fs41550-018-0406-3}{Nature Astronomy}, 2, 176

\bibitem[{Cicone {et~al.}(2014)Cicone, Maiolino, Sturm, Graci{\'{a} }-Carpio,
  Feruglio, Neri, Aalto, Davies, Fiore, Fischer, Garc{\'{\i}}a-Burillo,
  Gonz{\'{a}}lez-Alfonso, Hailey-Dunsheath, Piconcelli, \&
  Veilleux}]{Cicone2014}
Cicone, C., Maiolino, R., Sturm, E., {et~al.} 2014,
  \href{https://doi.org/10.1051%2F0004-6361%2F201322464}{A\&A}, 562, A21

\bibitem[{Crenshaw {et~al.}(2015)Crenshaw, Fischer, Kraemer, \&
  Schmitt}]{Crenshaw2015}
Crenshaw, D.~M., Fischer, T.~C., Kraemer, S.~B., \& Schmitt, H.~R. 2015,
  \href{https://doi.org/10.1088%2F0004-637x%2F799%2F1%2F83}{The Astrophysical
  Journal}, 799, 83

\bibitem[{Crenshaw \& Kraemer(2000)}]{Crenshaw2000b}
Crenshaw, D.~M. \& Kraemer, S.~B. 2000,
  \href{https://ui.adsabs.harvard.edu/abs/2000ApJ...532L.101C/abstract}{ApJ},
  532, 101

\bibitem[{Crenshaw {et~al.}(2000)Crenshaw, Kraemer, Hutchings, II, Gull,
  Kaiser, Nelson, Ruiz, \& Weistrop}]{Crenshaw2000a}
Crenshaw, D.~M., Kraemer, S.~B., Hutchings, J.~B., {et~al.} 2000,
  \href{https://doi.org/10.1086%2F301574}{AAS}, 120, 1731

\bibitem[{Cresci {et~al.}(2015{\natexlab{a}})Cresci, Mainieri, Brusa, \&
  et~al.}]{Cresci2015a}
Cresci, G., Mainieri, V., Brusa, M., \& et~al. 2015{\natexlab{a}},
  \href{https://doi.org/10.1088/0004-637X/799/1/82}{ApJ}, 799, 82

\bibitem[{Cresci \& Maiolino(2018)}]{Cresci2018}
Cresci, G. \& Maiolino, R. 2018,
  \href{https://doi.org/10.1038%2Fs41550-018-0404-5}{Nature Astronomy}, 2, 179

\bibitem[{Cresci {et~al.}(2015{\natexlab{b}})Cresci, Marconi, Zibetti,
  Risaliti, Carniani, Mannucci, Gallazzi, Maiolino, Balmaverde, Brusa, Capetti,
  Cicone, Feruglio, Bland-Hawthorn, Nagao, Oliva, Salvato, Sani, Tozzi,
  Urrutia, \& Venturi}]{Cresci2015b}
Cresci, G., Marconi, A., Zibetti, S., {et~al.} 2015{\natexlab{b}},
  \href{https://doi.org/10.1051%2F0004-6361%2F201526581}{A\&A}, 582, A63

\bibitem[{Cresci {et~al.}(2023)Cresci, Tozzi, Perna, Brusa, Marconcini,
  {Marconi, A.}, {Carniani, S.}, {Brienza, M.}, {Giroletti, M.}, {Belfiore,
  F.}, {Ginolfi, M.}, {Mannucci, F.}, {Ulivi, L.}, {Scholtz, J.}, {Venturi,
  G.}, {Arribas, S.}, {\"Ubler, H.}, {D\'{}Eugenio, F.}, {Mingozzi, M.},
  {Balmaverde, B.}, {Capetti, A.}, {Parlanti, E.}, \& {Zana, T.}}]{cresci2023}
Cresci, G., Tozzi, G., Perna, M., {et~al.} 2023,
  \href{https://doi.org/10.1051/0004-6361/202346001}{A\&A}, 672, A128

\bibitem[{Cresci {et~al.}(2017)Cresci, Vanzi, Telles, Lanzuisi, Brusa,
  Mingozzi, Sauvage, \& Johnson}]{Cresci2017}
Cresci, G., Vanzi, L., Telles, E., {et~al.} 2017,
  \href{https://doi.org/10.1051%2F0004-6361%2F201730876}{A\&A}, 604, A101

\bibitem[{Das {et~al.}(2005)Das, Crenshaw, Hutchings, Deo, Kraemer, Gull,
  Kaiser, Nelson, \& Weistrop}]{Das2005}
Das, V., Crenshaw, D.~M., Hutchings, J.~B., {et~al.} 2005,
  \href{https://doi.org/10.1086%2F432255}{AAS}, 130, 945

\bibitem[{Davies {et~al.}(2020)Davies, Baron, Shimizu, \& et~al.}]{Davies2020a}
Davies, R., Baron, D., Shimizu, T., \& et~al. 2020,
  \href{https://doi.org/10.1093/mnras/staa2413}{MNRAS}, 498, 4150–4177

\bibitem[{Di~Teodoro \& Fraternali(2015)}]{diteo2015}
Di~Teodoro, E. \& Fraternali, F. 2015,
  \href{https://arxiv.org/abs/1505.07834v1}{MNRAS}, 589, 3021

\bibitem[{Fabian(2012)}]{fabian2012}
Fabian, A.~C. 2012, \href{https://doi.org/10.48550/arXiv.1204.4114}{A\&A}, 50,
  455

\bibitem[{Ferrarese \& Ford(2005)}]{Ferrarese2005}
Ferrarese, L. \& Ford, H. 2005,
  \href{https://doi.org/10.1007%2Fs11214-005-3947-6}{Space Science Reviews},
  116, 523

\bibitem[{Feruglio {et~al.}(2015)Feruglio, Fiore, Carniani, Piconcelli,
  Zappacosta, Bongiorno, Cicone, Maiolino, Marconi, Menci, Puccetti, \&
  Veilleux}]{Feruglio2015}
Feruglio, C., Fiore, F., Carniani, S., {et~al.} 2015,
  \href{https://doi.org/10.1051%2F0004-6361%2F201526020}{A\&A}, 583, A99

\bibitem[{Feruglio(2010)}]{Feruglio2010}
Feruglio, C. e.~a. 2010,
  \href{https://doi.org/10.1051/0004-6361/201015164}{A\&A}, 518, L155

\bibitem[{Finlez {et~al.}(2022)Finlez, Treister, Bauer, Keel, Koss, Nagar,
  Sartori, Maksym, Venturi, Tub{\'{\i} }n, \& Harvey}]{Finlez2022}
Finlez, C., Treister, E., Bauer, F., {et~al.} 2022,
  \href{https://doi.org/10.3847%2F1538-4357%2Fac854e}{The Astrophysical
  Journal}, 936, 88

\bibitem[{Fiore {et~al.}(2017)Fiore, Feruglio, Shankar, Bischetti, Bongiorno,
  Brusa, Carniani, Cicone, Duras, Lamastra, Mainieri, Marconi, Menci, Maiolino,
  Piconcelli, Vietri, \& Zappacosta}]{Fiore2017}
Fiore, F., Feruglio, C., Shankar, F., {et~al.} 2017,
  \href{https://doi.org/10.1051%2F0004-6361%2F201629478}{A\&A}, 601, A143

\bibitem[{Fischer {et~al.}(2010)Fischer, Crenshaw, Kraemer, Schmitt, \&
  Trippe}]{Fischer2010}
Fischer, T.~C., Crenshaw, D.~M., Kraemer, S.~B., Schmitt, H.~R., \& Trippe,
  M.~L. 2010, \href{https://doi.org/10.1088%2F0004-6256%2F140%2F2%2F577}{AJ},
  140, 577

\bibitem[{{Fluetsch} {et~al.}(2019){Fluetsch}, {Maiolino}, {Carniani},
  {Marconi}, {Cicone}, {Bourne}, {Costa}, {Fabian}, {Ishibashi}, \&
  {Venturi}}]{Fluetsch2019}
{Fluetsch}, A., {Maiolino}, R., {Carniani}, S., {et~al.} 2019,
  \href{https://ui.adsabs.harvard.edu/abs/2019MNRAS.483.4586F}{\mnras}, 483,
  4586

\bibitem[{Fonseca-Faria {et~al.}(2021)Fonseca-Faria, Rodr{\'{\i} }guez-Ardila,
  Contini, \& Reynaldi}]{Fonseca2021}
Fonseca-Faria, M.~A., Rodr{\'{\i} }guez-Ardila, A., Contini, M., \& Reynaldi,
  V. 2021, \href{https://doi.org/10.1093%2Fmnras%2Fstab1806}{MNRAS}, 506, 3831

\bibitem[{Gagne {et~al.}(2014)Gagne, Crenshaw, Kraemer, Schmitt, Keel, Rafter,
  Fischer, Bennert, \& Schawinski}]{Gagne2014}
Gagne, J.~P., Crenshaw, D.~M., Kraemer, S.~B., {et~al.} 2014,
  \href{https://doi.org/10.1088%2F0004-637x%2F792%2F1%2F72}{The Astrophysical
  Journal}, 792, 72

\bibitem[{Gebhardt {et~al.}(2000)Gebhardt, Bender, Bower, Dressler, Faber,
  Filippenko, Green, Grillmair, Ho, Kormendy, Lauer, Magorrian, Pinkney,
  Richstone, \& Tremaine}]{Gebhardt2000}
Gebhardt, K., Bender, R., Bower, G., {et~al.} 2000,
  \href{https://doi.org/10.1086%2F312840}{ApJ}, 539, L13

\bibitem[{Genzel {et~al.}(2011)Genzel, Newman, Jones, Schreiber, Shapiro,
  Genel, Lilly, Renzini, Tacconi, Bouch{\'{e} }, Burkert, Cresci, Buschkamp,
  Carollo, Ceverino, Davies, Dekel, Eisenhauer, Hicks, Kurk, Lutz, Mancini,
  Naab, Peng, Sternberg, Vergani, \& Zamorani}]{Genzel2011}
Genzel, R., Newman, S., Jones, T., {et~al.} 2011,
  \href{https://doi.org/10.1088%2F0004-637x%2F733%2F2%2F101}{ApJ}, 733, 101

\bibitem[{Harrison {et~al.}(2018)Harrison, Costa, Tadhunter, Flütsch, Kakkad,
  Perna, \& Vietri}]{harrison2018}
Harrison, C.~M., Costa, T., Tadhunter, C.~N., {et~al.} 2018,
  \href{https://doi.org/10.1038%2Fs41550-018-0403-6}{Nature Astronomy}, 2, 198

\bibitem[{Hopkins \& Elvis(2010)}]{Hopkins2010}
Hopkins, P. \& Elvis, M. 2010,
  \href{https://www.semanticscholar.org/paper/Quasar-feedback%3A-more-bang-for-your-buck-Hopkins-Elvis/13b18cc94a0664c24ccfc942d39aaf4b431a4f38}{MNRAS},
  401, 7

\bibitem[{Juneau {et~al.}(2022)Juneau, Goulding, Banfield, Bianchi, Duc, Ho,
  Dopita, Scharwächter, Bauer, Groves, Alexander, Davies, Elbaz, Freeland,
  Hampton, Kewley, Nikutta, Shastri, Shu, Vogt, Wang, Wong, \&
  Woo}]{Juneau2022}
Juneau, S., Goulding, A.~D., Banfield, J., {et~al.} 2022,
  \href{https://doi.org/10.3847%2F1538-4357%2Fac425f}{ApJ}, 925, 203

\bibitem[{Karouzos {et~al.}(2016)Karouzos, Woo, \& Bae}]{Karouzos2016}
Karouzos, M., Woo, J.-H., \& Bae, H.-J. 2016,
  \href{https://doi.org/10.3847%2F0004-637x%2F819%2F2%2F148}{ApJ}, 819, 148

\bibitem[{Keel {et~al.}(2017)Keel, Lintott, Maksym, Bennert, Chojnowski,
  Moiseev, Smirnova, Schawinski, Sartori, Urry, Pancoast, Schirmer, Scott,
  Showley, \& Flatland}]{Keel2017}
Keel, W.~C., Lintott, C.~J., Maksym, W.~P., {et~al.} 2017,
  \href{https://doi.org/10.3847%2F1538-4357%2F835%2F2%2F256}{The Astrophysical
  Journal}, 835, 256

\bibitem[{Keel {et~al.}(2012)Keel, Lintott, Schawinski, Bennert, Thomas,
  Manning, Chojnowski, van Arkel, \& Lynn}]{Keel2012}
Keel, W.~C., Lintott, C.~J., Schawinski, K., {et~al.} 2012,
  \href{https://doi.org/10.1088%2F0004-6256%2F144%2F2%2F66} {The Astronomical
  Journal}, 144, 66

\bibitem[{Keel {et~al.}(2015)Keel, Maksym, Bennert, Lintott, Chojnowski,
  Moiseev, Smirnova, Schawinski, Urry, Evans, Pancoast, Scott, Showley, \&
  Flatland}]{Keel2015}
Keel, W.~C., Maksym, W.~P., Bennert, V.~N., {et~al.} 2015,
  \href{https://doi.org/10.1088%2F0004-6256%2F149%2F5%2F155}{The Astronomical
  Journal}, 149, 155

\bibitem[{{Kewley} {et~al.}(2019){Kewley}, {Nicholls}, {Sutherland}, {Rigby},
  {Acharya}, {Dopita}, \& {Bayliss}}]{Kewley2019}
{Kewley}, L.~J., {Nicholls}, D.~C., {Sutherland}, R., {et~al.} 2019,
  \href{https://ui.adsabs.harvard.edu/abs/2019ApJ...880...16K}{\apj}, 880, 16

\bibitem[{King(2003)}]{King2003}
King, A. 2003, \href{https://doi.org/10.48550/arXiv.astro-ph/0308342}{ApJ},
  149, 8

\bibitem[{King {et~al.}(2011)King, Zubovas, \& Power}]{King2011}
King, A.~R., Zubovas, K., \& Power, C. 2011,
  \href{https://doi.org/10.1111%2Fj.1745-3933.2011.01067.x}{MNRASL}, 415, L6

\bibitem[{Kraemer {et~al.}(2020)Kraemer, Turner, Couto, Crenshaw, Schmitt,
  Revalski, \& Fischer}]{Kraemer2020}
Kraemer, S.~B., Turner, T.~J., Couto, J.~D., {et~al.} 2020,
  \href{https://doi.org/10.1093%2Fmnras%2Fstaa428}{MNRAS}, 493, 3893

\bibitem[{Kudritzki {et~al.}(2021)Kudritzki, Teklu, Schulze, Remus, Dolag,
  Burkert, \& Zahid}]{Kudritzki2021}
Kudritzki, R.~P., Teklu, A.~F., Schulze, F., {et~al.} 2021,
  \href{https://iopscience.iop.org/article/10.3847/1538-4357/ac32cf}{AAS}, 910,
  87

\bibitem[{Marasco {et~al.}(2020)Marasco, Cresci, Nardini, Mannucci, Marconi,
  Tozzi, Tozzi, Amiri, Venturi, Piconcelli, Lanzuisi, Tombesi, Mingozzi, Perna,
  Carniani, Brusa, \& di~Serego~Alighieri}]{Marasco2020}
Marasco, A., Cresci, G., Nardini, E., {et~al.} 2020,
  \href{https://doi.org/10.1051%2F0004-6361%2F202038889}{A\&A}, 644, A15

\bibitem[{Marconi \& Hunt(2003)}]{Marconi2003}
Marconi, A. \& Hunt, L. 2003,
  \href{https://doi.org/10.48550/arXiv.astro-ph/0304274}{ApJ}, 589, 21

\bibitem[{Mingozzi {et~al.}(2019)Mingozzi, Cresci, Venturi, Marconi, Mannucci,
  Perna, Belfiore, Carniani, Balmaverde, Brusa, Cicone, Feruglio, Gallazzi,
  Mainieri, Maiolino, Nagao, Nardini, Sani, Tozzi, \& Zibetti}]{Mingozzi2018}
Mingozzi, M., Cresci, G., Venturi, G., {et~al.} 2019,
  \href{https://doi.org/10.1051%2F0004-6361%2F201834372}{A\&A}, 622, A146

\bibitem[{Morganti(2017)}]{morganti2017}
Morganti, R. 2017, \href{https://doi.org/10.3389/fspas.2017.00042}{Frontiers},
  4, 87

\bibitem[{Müller-S{\'{a} }nchez {et~al.}(2011)Müller-S{\'{a} }nchez, Prieto,
  Hicks, Vives-Arias, Davies, Malkan, Tacconi, \& Genzel}]{MullerSanchez2011}
Müller-S{\'{a} }nchez, F., Prieto, M.~A., Hicks, E. K.~S., {et~al.} 2011,
  \href{https://doi.org/10.1088%2F0004-637x%2F739%2F2%2F69}{ApJ}, 739, 69

\bibitem[{Nardini \& Zubovas(2018)}]{Nardini2018}
Nardini, E. \& Zubovas, K. 2018,
  \href{https://doi.org/10.1093/mnras/sty1144}{MNRAS}, 478, 2274–2280

\bibitem[{Oh {et~al.}(2008)Oh, de~Blok, Walter, Brinks, \& Kennicutt}]{oh2008}
Oh, S.-H., de~Blok, W. J.~G., Walter, F., Brinks, E., \& Kennicutt, R.~C. 2008,
  \href{https://doi.org/10.1088%2F0004-6256%2F136%2F6%2F2761}{A\&A}, 136, 2761

\bibitem[{{Osterbrock}(1989)}]{Osterbrock2006}
{Osterbrock}, D.~E. 1989, {Astrophysics of gaseous nebulae and active galactic
  nuclei}

\bibitem[{Perna {et~al.}(2017)Perna, Lanzuisi, Brusa, Mignoli, \&
  Cresci}]{Perna2017}
Perna, M., Lanzuisi, G., Brusa, M., Mignoli, M., \& Cresci, G. 2017,
  \href{https://doi.org/10.1051%2F0004-6361%2F201630369}{A\&A}, 603, A99

\bibitem[{Richings \& Faucher-Giguère(2018)}]{Richings2018}
Richings, A.~J. \& Faucher-Giguère, C.~A. 2018,
  \href{https://doi.org/10.1093/mnras/sty1285}{MNRAS}, 478, 2274–2280

\bibitem[{Rupke {et~al.}(2005)Rupke, Veilleux, \& Sanders}]{Rupke2005}
Rupke, D.~S., Veilleux, S., \& Sanders, D.~B. 2005,
  \href{https://doi.org/10.1086%2F432886}{ApJS}, 160, 87

\bibitem[{Schmitt \& Kinney(2016)}]{Schmitt2016}
Schmitt, H. \& Kinney, A. 2016,
  \href{https://ui.adsabs.harvard.edu/abs/1996ApJ...463..498S/abstract}{MNRAS},
  463, 498

\bibitem[{Sofue(2015)}]{Sofue_2015}
Sofue, Y. 2015, \href{https://doi.org/10.1093/pasj/psv042}{Publications of the
  Astronomical Society of Japan}, 67, 75

\bibitem[{Storchi-Bergmann {et~al.}(2010)Storchi-Bergmann, Lopes, McGregor,
  Riffel, Beck, \& Martini}]{Storchi2010}
Storchi-Bergmann, T., Lopes, R. D.~S., McGregor, P.~J., {et~al.} 2010,
  \href{https://doi.org/10.1111%2Fj.1365-2966.2009.15962.x}{MNRAS}, 402, 819

\bibitem[{Sturm(2011)}]{Sturm2011}
Sturm, E. e.~a. 2011, \href{https://doi.org/10.1088/2041-8205/733/1/L16}{ApJ},
  733, 16

\bibitem[{Tombesi {et~al.}(2015)Tombesi, Mel{\'{e}}ndez, Veilleux, Reeves,
  Gonz{\'{a}}lez-Alfonso, \& Reynolds}]{Tombesi2015}
Tombesi, F., Mel{\'{e}}ndez, M., Veilleux, S., {et~al.} 2015,
  \href{https://doi.org/10.1038%2Fnature14261}{Nature}, 519, 436

\bibitem[{Tozzi {et~al.}(2021)Tozzi, Cresci, Marasco, Nardini, Marconi,
  Mannucci, Chartas, Rizzo, Amiri, Brusa, Comastri, Dadina, Lanzuisi, Mainieri,
  Mingozzi, Perna, Venturi, \& Vignali}]{Tozzi2021}
Tozzi, G., Cresci, G., Marasco, A., {et~al.} 2021,
  \href{https://doi.org/10.1051%2F0004-6361%2F202040190}{A\&A}, 648, A99

\bibitem[{Urry \& Padovani(1995)}]{UrryPadovani1995}
Urry, C.~M. \& Padovani, P. 1995,
  \href{https://ui.adsabs.harvard.edu/abs/1995PASP..107..803U/abstract}{Publications
  of the Astronomical Society of the Pacific}, 107, 803

\bibitem[{Veilleux {et~al.}(2017)Veilleux, Bolatto, Tombesi, Mel{\'{e} }ndez,
  Sturm, Gonz{\'{a}}lez-Alfonso, Fischer, \& Rupke}]{veilleux2017}
Veilleux, S., Bolatto, A., Tombesi, F., {et~al.} 2017,
  \href{https://doi.org/10.3847%2F1538-4357%2Faa767d}{ApJ}, 843, 18

\bibitem[{Veilleux {et~al.}(2018)Veilleux, Maiolino, Bolatto, \&
  Aalto}]{Kakkad2018}
Veilleux, S., Maiolino, R., Bolatto, A.~D., \& Aalto, S. 2018,
  \href{https://doi.org/10.1051/0004-6361/201832790}{A\&A}, 618, A6

\bibitem[{Veilleux {et~al.}(2020)Veilleux, Maiolino, Bolatto, \&
  Aalto}]{Veilleux2020}
Veilleux, S., Maiolino, R., Bolatto, A.~D., \& Aalto, S. 2020,
  \href{https://doi.org/10.1007%2Fs00159-019-0121-9}{A\&AReview}, 28

\bibitem[{Venturi {et~al.}(2021)Venturi, Cresci, Marconi, Mingozzi, Nardini,
  Carniani, Mannucci, Marasco, Maiolino, Perna, Treister, Bland-Hawthorn, \&
  Gallimore}]{Venturi2021}
Venturi, G., Cresci, G., Marconi, A., {et~al.} 2021,
  \href{https://doi.org/10.1051%2F0004-6361%2F202039869}{A{\&}A}, 648, A17

\bibitem[{Venturi \& Marconi(2020)}]{Venturi2020}
Venturi, G. \& Marconi, A. 2020,
  \href{https://doi.org/10.1017/S1743921320002203}{PIAUS}, 359, 8

\bibitem[{Venturi {et~al.}(2017)Venturi, Marconi, Mingozzi, Carniani, Cresci,
  Risaliti, \& Mannucci}]{Venturi2017}
Venturi, G., Marconi, A., Mingozzi, M., {et~al.} 2017,
  \href{https://doi.org/10.3389%2Ffspas.2017.00046}{Frontiers in Astronomy and
  Space Sciences}, 4

\bibitem[{{Venturi} {et~al.}(2018){Venturi}, {Nardini}, {Marconi}, {Carniani},
  {Mingozzi}, {Cresci}, {Mannucci}, {Risaliti}, {Maiolino}, {Balmaverde},
  {Bongiorno}, {Brusa}, {Capetti}, {Cicone}, {Ciroi}, {Feruglio}, {Fiore},
  {Gallazzi}, {La Franca}, {Mainieri}, {Matsuoka}, {Nagao}, {Perna},
  {Piconcelli}, {Sani}, {Tozzi}, \& {Zibetti}}]{Venturi2018}
{Venturi}, G., {Nardini}, E., {Marconi}, A., {et~al.} 2018,
  \href{https://ui.adsabs.harvard.edu/abs/2018A&A...619A..74V}{\aap}, 619, A74

\bibitem[{Vukcevic(2021)}]{Vuckcevic2021}
Vukcevic, M. 2021,
  \href{https://iopscience.iop.org/article/10.3847/1538-3881/abd568}{ApJ}, 161,
  118

\bibitem[{Yoon {et~al.}(2021)Yoon, Park, Chung, \& Zhang}]{Yoon2021}
Yoon, Y., Park, C., Chung, H., \& Zhang, K. 2021,
  \href{https://arxiv.org/abs/2110.06033}{AAS}, 922, 249

\bibitem[{Zakamska \& Greene(2014)}]{Zakamska2014}
Zakamska, N.~L. \& Greene, J.~E. 2014,
  \href{https://doi.org/10.1093/mnras/stu842}{MNRAS}, 442, 784

\bibitem[{Zubovas \& King(2012)}]{Zubovas2012}
Zubovas, K. \& King, A. 2012,
  \href{https://doi.org/10.48550/arXiv.1201.3540}{ApJ}, 460, 235

\end{thebibliography}

\end{document}